\documentclass[fleqn,usenatbib,useAMS]{mnras}

\usepackage[T1]{fontenc}
\usepackage{ae,aecompl}

\usepackage{graphicx}	
\usepackage{amsmath}	
\usepackage{amssymb}	
\usepackage{pdfpages}

\newcommand{\be}{\begin{equation}}
\newcommand{\ee}{\end{equation}} 
\newcommand{\bse}{\begin{subequations}}
\newcommand{\ese}{\end{subequations}} 
\newcommand{\bary}{\begin{eqnarray}}
\newcommand{\eary}{\end{eqnarray}}

\newcommand{\epsr}{\rm \epsilon_{\rm r}}
\newcommand{\mbh}{ M_{\rm BH}}

\newcommand{\mdotbh}{\dot{M}_{\rm BH}}

\newcommand{\msun}{\rm M_{\odot}}
 
\newcommand{\mdotbondi}{\dot{M}_{\rm Bondi}} 
\newcommand{\mdotedd}{\dot{M}_{\rm Edd}} 
\newcommand{\mstar}{M_{\rm star}}
\newcommand{\K}{\rm K}

\newcommand{\mcrit}{ M_{200}}

\newcommand{\Ledd}{\lambda_{\rm Edd}}
\newcommand{\LhX}{L_{\rm HX}}
\newcommand{\Lbol}{L_{\rm bol}}
\newcommand{\ergs}{\mbox{erg}~\mbox{s}^{-1}}
\newcommand{\kpc}{\rm kpc}
\newcommand{\cMpc}{\rm cMpc}
\newcommand{\Mpc}{\rm Mpc}
\newcommand{\erg}{\rm erg}
\newcommand{\lsim}{\mathrel{\hbox{\rlap{\lower.55ex\hbox{$\sim$}} \kern-.3em\raise.4ex\hbox{$<$}}}}
\newcommand{\gsim}{\mathrel{\hbox{\rlap{\lower.55ex\hbox{$\sim$}} \kern-.3em\raise.4ex\hbox{$>$}}}}

\newcommand{\LsX}{L_{\rm SX}}
\newcommand{\REF}{Ref-L100N1504}

\newcommand{\REFFIFTY}{Ref-L050N0752}
\newcommand{\AGNdT}{AGNdT9-L050N0752}
\newcommand{\RECAL}{Recal-L025N0752}
\newcommand{\SmallSeeds}{Small-seeds-L050N0752}

\defcitealias{schaye2015}{S15}
\defcitealias{crain2015}{C15}

\title[ SMBHs in EAGLE Universe]{ Supermassive black holes in the EAGLE Universe. Revealing the observables 
of their growth.}

\author[Rosas-Guevara et al.]{
Yetli Rosas-Guevara $^{1,2}$\thanks{E-mail: y.rosas@das.uchile.cl}; 
Richard G. Bower $^{2}$ \thanks{E-mail: r.g.bower@dur.ac.uk};
Joop Schaye $^3$; 
Stuart McAlpine $^2$; 
\newauthor
Claudio Dalla Vecchia $^{4,5}$; 
Carlos S. Frenk $^2$;
Matthieu Schaller $^2$;
Tom Theuns $^2$. \\
$^{1}$DAS, University of Chile, Camino del Observatorio 1515, Las Condes, Santiago, Chile.\\
$^{2}$ICC, Physics Department, University of Durham, South Road, Durham DH1 3LE, UK \\
$^{3}$Leiden Observatory, Leiden University, P.O. Box 9513, 2300 RA Leiden, The Netherlands\\
$^{4}$Instituto de Astrof\'isica de Canarias, C/ V\'ia L\'actea s/n, 38205 La Laguna, Tenerife, Spain \\
$^{5}$Departamento de Astrof\'isica, Universidad de La Laguna, Av.~del Astrof\'isico Francisco S\'anchez s/n, 
38206 La Laguna, Tenerife, Spain}

\date{Accepted XXX. Received YYY; in original form ZZZ}

\pubyear{2016}

\begin{document}
\label{firstpage}
\pagerange{\pageref{firstpage}--\pageref{lastpage}}
\maketitle

\begin{abstract}
We investigate the evolution of supermassive black holes in the `Evolution and Assembly
of GaLaxies and their Environments' (EAGLE) cosmological hydrodynamic
simulations.  The largest of the EAGLE volumes covers a
$(100 \,\cMpc)^3$ and includes state-of-the-art physical models for star formation
and black hole growth that depend only on local gas properties. We focus on the black hole mass function, 
Eddington ratio distribution and the implied duty cycle of nuclear activity. 
The simulation is broadly consistent with observational constraints on these quantities. 
In order to make a more direct comparison with observational data, we calculate the soft and hard X-ray
luminosity functions of the active galactic nuclei (AGN). Between redshifts $0$  and $1$, the simulation is in agreement with 
data. At higher redshifts, the simulation tends to underpredict the luminosities of
the brightest observed AGN. This may be due to the limited volume of
the simulation, or a fundamental deficiency of the underlying model.
It seems unlikely that additional unresolved variability can account for this difference.
The simulation shows a similar `downsizing' of the AGN population as seen in observational surveys. 
\end{abstract}

\begin{keywords}
black hole physics, galaxies: formation, galaxies: active, 
methods: hydrodynamical simulations, quasars: general.
\end{keywords}



\section{Introduction}

Accreting supermassive black holes (SMBHs) are one of the most efficient sources of radiation in
the Universe. During periods of strong activity, they are often prominent as optical nuclei of galaxies, and
referred to as Active Galactic Nuclei (AGN). From a theoretical perspective, the vast energy outputs
of AGN offer an appealing explanation for 
the steep cut-off of the massive end of the galaxy luminosity function (e.g.
\citealt{bower2006}, \citealt{croton2006}) and the scaling of the
X-ray properties of galaxy groups and clusters
(e.g. \citealt{binney_tabor1995}, \citealt{churazov2001}, \citealt{mcCarthy2010}). 

From an observational  perspective, the strong correlation between
the mass of the central SMBH and the properties of the host galaxy, such as
 its velocity dispersion and bulge mass (see review by \citealt{kormendy_ho2013}, also by \citealt{graham2016}), is consistent with a causal connection. 
One way to explore this connection, is to examine the
evolution of AGN, for example by constructing luminosity functions at different cosmic epochs. 
Integrating the total energy radiated over the AGN lifetime then provides a method of charting the build-up 
of the rest-mass energy of SMBHs \citep{soltan1982}. 

Measurements of the luminosity distribution of AGN require large, unbiased samples selected over a wide range of redshifts and luminosities. Constructing such samples is difficult because a fraction of the emission that emerges from the SMBH is obscured by the surrounding gas and dust making  an uncertain fraction of SMBH difficult to detect \citep{lansbury2015}.  
Although spectroscopic optical surveys are able to scan wide areas and detect large numbers of AGN up to redshift $z=6$, 
these surveys are biased to the brightest and most unobscured population of SMBHs.  While the
mid-infrared band can also be used to detect SMBHs via the reprocessed
emission from dust heated by AGN activity, the emission from the SMBH is often overwhelmed by 
the  host galaxy. X-rays therefore provide the most efficient and unbiased method of selection.
Although soft X-rays are the most easily observed band, AGN selection is still biased due to gas extinction around the SMBH. 
This makes hard X-rays the least biased wavelength range  to detect the full SMBH population as obscuration is
greatly reduced. Recently, multiple large X-ray surveys have been carried out by \cite*{ueda2014}, \cite*{aird2015} and \cite*{buchner2015}. These studies have revealed that the AGN population evolves strongly and that their number density 
abruptly decreases between $z\approx (1-2)$ and today. Moreover, these studies show that there is strong `downsizing' of the AGN
population in the sense that the space density of higher-luminosity AGN peaks at higher
redshifts.

Such deep X-ray surveys provide tests for models linking the build up of 
galaxies and their SMBHs. Recently, this has been explored using semi-analytic models where the
growth of SMBHs and AGN feedback have been incorporated as analytic
approximations. Typically, these models assume that AGN activity is triggered  by major galaxy
mergers or disc instabilities, and calibrate AGN feedback to reproduce the galaxy mass function
 \citep{bower2006,croton2006,delucia2007}. After accounting for 
strong dust obscuration of faint AGN, such models have been able to reproduce  
the observed AGN luminosity functions and AGN 
downsizing \citep{fanidakis12,hirschmann2012}.  Such studies
rely on summarising complex hydrodynamic interactions by simple models, but provide important and useful approximations.

Hydrodynamic simulations offer an alternative approach, more clearly differentiating the resolved hydrodynamical interactions from the small-scale processes that cannot be directly resolved.
AGN evolution has been explored by hydrodynamical simulations of
isolated galaxy mergers \citep{springel2005a} and small cosmological volumes at high redshift. In these simulations, SMBH growth and AGN feedback are incorporated as subgrid physics
\citep{springel2005a, booth_schaye2009}. SMBH accretion is typically based on the pure
Bondi-Hoyle model \citep{bondi44} or on simple modifications of this
\citep{springel2005a,booth_schaye2009}. Recent studies, however, have recognised 
the importance of accounting for the effects of accreting high angular momentum material (e.g. \citealt{hopkins2009,rosas-guevara2015,angles-alcazar2016}). 

In addition, a fraction of the rest-mass energy of the accretion flow may be injected in the surrounding gas as thermal energy or a momentum driven wind or jet.
Since such processes cannot be directly resolved, simulations choose to implement feedback in different ways, for example as thermal heating proportional to the mass accretion rate (e.g. \citealt{springel2005a}, \citealt{booth_schaye2009}), by explicitly distinguishing quasar and radio modes (e.g.
\citealt{sijacki2007}, \citealt{vogelsberger2014}), or by  injection of momentum into the surrounding gas
(e.g. \citealt{power2011, choi2013}).     

The latest generation of cosmological hydrodynamic simulations can now track the
evolution of a galaxy  population resolving the formation of individual
galaxies with a resolution of $\sim 700 \, \rm pc$ within large cosmological volumes, typically $100 \, \cMpc$ on a side \citep{vogelsberger2014,schaye2015,khandai2015}. In this paper 
we will focus on the Evolution and Assembly of GaLaxies
and their Environments (EAGLE) simulations \citep{schaye2015,crain2015}.  The EAGLE simulations reproduce  many properties of galaxies, such as: the evolution of the galaxy mass functions \citep{furlong2015a}, the evolution 
of galaxy sizes \citep{furlong2015b}, the colour-magnitude diagram \citep{trayford2016} and the properties of molecular and atomic gas \citep{lagos2015,bahe2016}.
Much of the  success  in reproducing the properties of the massive galaxies is due to
the effects of AGN feedback \citep{crain2015}.  A key question is, therefore, whether the model
reproduces  the observables of SMBH evolution in a cosmological context.  This test is almost entirely independent
of the calibration procedure used to select model parameters, since the calibration procedure only considered the normalisation of the correlation between the present-day SMBH mass and galaxy stellar mass \citep{schaye2015}.

In this paper we present the evolution of SMBHs in the EAGLE simulations from $z=11$ to $0$. We compare  the predicted X-ray observables in EAGLE to  observational data from X-ray deep fields up to redshift 7.
Such deep fields roughly correspond to the size of the largest EAGLE simulation, implying that we are restricted  
to densities of moderately luminous AGN.  In section~\ref{sec:code_and_sim}, we briefly outline 
the relevant subgrid physics used in the EAGLE project and describe how we compute the intrinsic X-ray
emission from AGN using empirical corrections for the bolometric
luminosity and the obscured fraction.  In section~\ref{sec:results}, we present the results of the simulation. We summarize the properties of local SMBHs, such as their mass function in
section~\ref{sub:bhmf_z0}.  The evolution of the black hole mass function,
the Eddington ratio distribution plane and the black hole mass-halo mass relation 
are investigated in section~\ref{sub:highz-bh}.  In section \ref{sub:AGNlf} we compare
the evolution of the  AGN luminosity function in X-ray bands with the most
recent observational estimates.  We show that AGN in EAGLE follow a similar downsizing trend
to that seen in observational data.  Finally in section~\ref{sec:discussions}, we
summarise and  discuss our main results.  Additional tests of simulation convergence and parameter dependencies are  
given in Appendix~\ref{appendix:convergencetests} and in Appendix~\ref{appendix:choiceaccretionregime}.

\section{Code and Simulations}
\label{sec:code_and_sim}
\subsection{Code}
In this study we use simulations from the EAGLE project\footnote{http://eaglesim.org \\
http://eagle.strw.leidenuniv.nl}.  This  consists  of a large number of
cosmological simulations, with variations in parameters, galaxy formation
subgrid models and numerical resolutions, as well as a large, $(100 \cMpc)^3$ volume
reference calculation.  Full details of the EAGLE
simulations can be found in  \cite{schaye2015} and  \cite{crain2015} (hereafter
\citetalias{schaye2015} and \citetalias{crain2015}); here we give only a brief overview.  

The EAGLE simulations were performed with a modified version of the parallel
hydrodynamic code  Gadget-3 which is a computationally efficient version of the
public  code Gadget-2 \citep{springel2005b}. The improvements to the hydrodynamics solver, which are collectively referred to as  
Anarchy, aim to better model hydrodynamical instabilities,  as described in Dalla
Vecchia (in prep) (see also \citetalias{schaye2015} and \citealt{schaller2015}). Here,  we concentrate on
the reference model denoted  as, \REF, which corresponds to a cubic volume
of $L=100$ comoving  Mpc ($\rm cMpc$) on a side. Initially, it  employs
$2\times 1504^3$ particles. In order to study
numerical {\it weak } convergence, we also use the simulations \AGNdT~and \RECAL~with box sizes 
 $L=50$ and $25$ $\rm cMpc$ respectively, containing $2\times 752^3$ particles per simulation. Numerical {\it weak } convergence   is defined in
\citetalias{schaye2015}  and reflects the need of recalibrating the subgrid parameters
to model more faithfully  the physical processes at increasing resolution.  
Further simulation variations are considered in
Appendix~\ref{appendix:convergencetests}. 

The EAGLE simulations start from cosmological initial conditions
at $z=127$. The  transfer  function for the linear matter power
spectrum was generated with CAMB \citep{lewis2000}, adopting the Planck Cosmology parameters
\citep{planck13}.  The Gaussian initial conditions  were generated using the
linear matter power spectrum  and the random phases from the public multiscale
white noise Panphasia field \citep{jenkins2013}.  Particle displacements and
velocities are  calculated using second-order perturbation theory
\citep{jenkins2010}. 

The setup of these simulations gives  a mass resolution of  $9.7\times 10^6
\msun$ for dark matter ( and $1.81 \times 10^6 \msun$ for baryonic) particles.  The
gravitational interaction between particles is calculated using a Plummer
potential with a softening length of $2.66$ comoving $\kpc$ limited to a
maximum physical size  of $0.70$ $\kpc$.   The box sizes, particle numbers and
 mass and spaced resolutions are summarised in Table \ref{table:simulations}.

\begin{table*}
\caption{Box length, initial particle number, initial baryonic and dark matter particle mass, comoving and maximum proper gravitational softening 
for the EAGLE simulations used in this paper. }
\begin{tabular}{|l|l|l|c|c|c|c|}
\hline
Name& L               &   $N$             &     $m_{\rm g}$         &  $m_{\rm DM}$ &  $\epsilon_{\rm com}$ & $\epsilon_{\rm prop}$ \\
    & [$\rm cMpc$]    &                   & [$\msun$]           & [$\msun$] &     $[\rm ckpc]$ &          $[\rm pkpc]$ \\ 
\hline
Ref-L100N1504 & $100$  & $2\times1504^3$  &  $1.81\times 10^6$& $9.70\times10^6$&  $2.66$     &  $0.70$ \\                

AGNdT9-L050N0752 & $50$  & $2\times752^3$  &  $1.81\times 10^6$& $9.70\times10^6$&  $2.66$     &  $0.70$ \\                

Recal-L025N0752 & $25$  & $2\times752^3$  &  $2.26\times 10^5$& $1.21\times10^6$&  $1.33$     &  $0.35$ \\               
\hline
\end{tabular}
\label{table:simulations}
\end{table*}

\subsection{Subgrid physics}     
The galaxy formation subgrid physics included in these simulations is
largely based on that used for the OWLS project (\citealt{scha10}, see also
\citealt{crain09}). Many improvements have been implemented, in particular in
the modelling of stellar feedback and black hole growth. We provide a brief overview below.
Further details can be found in \citetalias{schaye2015} and  an extensive comparison of the effects of varying the
subgrid physics parameters is given in \citetalias{crain2015}. The
values of the parameters that differ between the simulations can be found in Table \ref{table:parameters}. 

\begin{itemize}

\item {\it Radiative cooling and photoheating, star formation and stellar feedback}. 

Radiative cooling and photoheating are as described in
\cite*{wiersma2009a}.  The radiative  rates are computed element by element  in
the presence of the cosmic microwave background (CMB) and the UV and X-ray
background radiation from quasars and galaxies (model of \citealt{haardt2001}).
Eleven elements are tracked. The radiative cooling and heating rates are computed
with the software Cloudy \citep{ferland2013}. Prior to reionization,
the gas is in collisional ionization equilibrium and no ionizing background is
present.

Star formation is
implemented following the model of \cite*{scha08}, including a metallicity dependent density threshold, $n^{*}_{\rm
z}\sim Z^{-0.64}$ \citep{schaye2004} above which gas particles are allowed to form stars. The model parameters
are chosen to reproduces the empirical
\citeauthor{schmidt1959}-\citeauthor{kennicut1998} law which is encoded in
terms of a pressure law. A temperature floor is
imposed as a function of density, $P \propto \rho^{\gamma\rm eff}$, for gas with
$\gamma_{\rm eff} = 4/3$. This value of $\gamma_{\rm eff}$ leads to the Jeans mass, and the ratio of the Jeans length
to the SPH kernel length, being independent of density, avoiding spurious
fragmentation due to a lack of resolution. Gas particles are  stochastically selected
for star formation and converted to collisionless star particles. Each star particle
represents a simple stellar population formed with  a \cite*{chabrier2003}
IMF. 

Stellar evolution is  implemented  as  described in \citetalias{schaye2015} and
\cite*{wiersma2009b}. The  stellar mass loss and consequent metal enrichment of 11
elements are  modelled via three channels: (1) AGB stars, (2) Supernova (SNe) type Ia and (3) Massive stars
and core collapse SNe. The mass loss of the stellar population, including metals, is added to the gas particles that are within an
SPH kernel  of the star particle. 

Feedback from star formation is  treated stochastically, using the thermal
injection method described in \cite*{dallavecchia_schaye12}. The total energy
available to inject into the ISM  per core SN  is assumed to be $10^{51}\erg$.
This amount of energy is injected 30 Myr
after the birth of the  star particle. Neighbouring gas particles are selected to
be  heated stochastically based on the available energy, and then heated by a fixed
temperature difference of  $\Delta T= 10^{7.5}\rm K$.  The stochastic heating
distributes  the energy such that the cooling time relative to the sound
crossing time across a resolution element allows the thermal energy to be
converted to kinetic energy, limiting spurious losses.  
The fraction of the available energy injected into the ISM depends on the
local gas metallicity and density.

\item{\it Black hole growth and AGN feedback}. 

Halos that become more massive than  $1.48 \times 10^{10} \msun$   
are seeded with black holes of $1.48 \times 10^5 \msun$ ($1.48 \times 10^4 \msun$ for the simulation
\SmallSeeds~ presented in Appendix~\ref{appendix:convergencetests})
using the method of \citet{springel2005a}. In order to mimic dynamical
friction, at each timestep the black holes less massive than 100 times the initial mass of the gas particles are
relocated to the minimum of its local gravitational potential.  
SMBHs can then grow via gas accretion, where the accretion rates are calculated by  the modified Bondi-Hoyle  
model presented in \citet{rosas-guevara2015}:
\begin{equation} \dot{M}_{\rm accr} = {\rm min}(\dot{M}_{\rm
Bondi}\left[ C^{-1}_{\rm visc}({ c}_{\rm s}/V_\Phi)^3 \right], \dot{M}_{\rm
Bondi}), 
\end{equation} 
where $c_{\rm s}$ is the sound speed, $V_\Phi$ is
the SPH-average circular speed of the gas around the black hole and $C_{\rm
visc}$  is a viscosity parameter  that controls  the degree of modulation of
the Bondi-Hoyle rate, $\mdotbondi$, in high circulation flows. SMBH accretion rates  are also 
Eddington limited. In contrast to \cite{rosas-guevara2015} and \cite{booth_schaye2009}, EAGLE accretion rates do not include an additional `$\beta$-factor' to boost the accretion rates when the surrounding gas density is high. This parameter is largely
degenerate with the $C_{\rm visc}$ parameter. The values of $C_{\rm visc}$ in the simulations are found in Table~\ref{table:parameters}.  SMBHs also grow via mergers when they are within their smoothing length and have sufficiently small radial velocity. Further details are given in \citetalias{schaye2015}.
Following \citet{springel2005a}, two masses are adopted for BH particles: a sub grid mass that is applied to the computation of the  gas accretion rates  and AGN feedback,  and a particle mass that is used in the gravitational calculations. Initially, the sub-grid mass is smaller than the particle mass. Once the subgrid mass exceeds the particle mass, the SMBH accretes stochastically  gas particles in its vicinity so both masses grow a step. This method ensures that sub-grid masses can be smaller than particle mass whilst conserving  the gravitating mass.

AGN feedback is implemented following the stochastic model of \cite*{booth_schaye2009}.  Thermal energy is injected into the surrounding gas as a fraction of the rest mass energy of the gas accreted by the SMBH.  Neighbouring gas particles of the SMBH are stochastically selected and heated by a
temperature difference of $\Delta T= 10^{8.5}\K$ for the simulation \REF~and $10^{9}\K$  for the simulation \AGNdT.    
The scheme is similar to that used to implement feedback from star formation, but uses a significantly higher heating temperature for the energy injection events. It is important to emphasise the simplicity of the feedback scheme that we adopt: a single {\it mode} of AGN feedback is 
implemented throughout using a fixed efficiency of $0.1$, from which, a fraction of 0.15 is coupled to the surrounding gas. 

\end{itemize}

\begin{table}
\centering
\caption{Values of parameters that differ between the simulations. These parameters affects the subgrid physics from
 star formation and from black holes used in this work; $n_{\rm H,0}$ y $n_n$ affect the fraction of 
the available energy injected from SN II into the ISM (see \citetalias{schaye2015});$C_{\rm visc}$ and $\Delta$T$_{\rm AGN}$ affect
the BH accretion rates and the energy released from AGN as indicated in the text.}
\smallskip
 \begin{minipage}{8.5cm}
\centering
  \begin{tabular}{|l|c|c|c|l|}
   \hline
   {\bf Simulation Prefix}  & $n_{\rm H,0}$ & $n_n$ & $C_{\rm visc}$ & $\Delta$T$_{\rm AGN}$ \\
   
                            &   [cm$^{-3}$]  &       &         & [K]  \\
   \hline
   Ref      		     &  0.67        &  2/ln 10     & 2$\pi$  & 10$^{8.5}$   \\
   AGNdT9   		     & 0.67         &  2/ln 10    & 2$\pi \times 10^2$   & 10$^{9}$     \\
   Recal    		     & 0.25         &  1/ln 10  & 2$\pi \times 10^3$   & 10$^{9}$    \\ 
   \hline
  \end{tabular}  \par
  \end{minipage}
\label{table:parameters}
\end{table}

\subsection{Simulation outputs}
\label{subsec:outputs}
Most published papers by the EAGLE collaboration are based on the analysis of 29 ``snapshot'' outputs, 
containing the full information on all particles at a particular redshift.  These provide
a good census of the masses of SMBHs at one particular time. Since we are interested in the dominant black hole
of each dark matter halo, we do not use the quantities tabulated in the public database \citep{mcAlpine2015}
as these correspond to summed quantities of all SMBHs within the halo. As discussed in that paper, these can differ 
significantly in the case of low-mass black holes. 

Although ``snapshot'' outputs can be used to construct the AGN luminosity function, the strong variability of 
AGN in the simulation means that the statistics of luminous AGN are poorly sampled. To obtain a better
determination of the AGN luminosity functions, we make use of the more frequent ``snipshot'' outputs. These are partial copies of the particle 
state of the simulation, which are output in order to track critical simulation quantities with higher time 
resolution. There are 406 (400, 406) `snipshots' outputs 
for the \REF~(\AGNdT, \RECAL) simulation, with a temporal separation between $10$ and  $60$ Myr. In the
simulation, AGN are highly variable on significantly shorter timescales, and we average the luminosity functions
in ranges of snipshots in order to improve the statistical sampling of luminous outbursts. A detailed analysis of AGN variability will be presented in McAlpine et al. (2016, in preparation).
Although this procedure allows us to reduce sampling uncertainties due to variability, it does 
not allow us to include rare objects that are not present in the simulation volume. In appendix \ref{appendix:convergencetests}, we compare 
simulations with different volumes, but the same parameters. The analysis presented then suggests that variability is the dominant uncertainty
and that the procedure we use does not appear to cause a significant underestimate of the abundance of luminous AGN.

To give an impression of the size of the SMBH population, the first SMBH
appears in the \REF~simulation at $z=14.5$, the most massive black hole has a
mass of $\mbh = 4.1\times 10^9 \msun$ and is located in the most massive halo (which has a mass of
$\mcrit=6.4\times 10^{14}\msun$)  at $z=0$. The total number of SMBHs at $z=0$ 
with mass larger than $10^6\msun$ is $5627$ of which $25$ have masses $>10^9\msun$,
$505$ have masses $> 10^{8}\msun$ and $1996$ have masses $>10^7\msun$.  

\subsection{Post-processing and definition of accretion regimes}

\label{sec:accretion-regimes}
As we have previously stressed, the subgrid models of SMBH accretion and feedback used do not make any distinction between 
different regimes of SMBH accretion. {\it In post-processing,} we distinguish between 
the activity levels of SMBHs based on their Eddington ratio, 
\be
\Ledd\equiv \dot{M}_{\rm accr}/\mdotedd,
\label{eq:edd_ratio}
\ee 
where $\dot{M}_{\rm accr}$ and $\mdotedd$ are the SMBH mass accretion rate and the Eddington
limit respectively.  We define two `active' accretion regimes. For  Eddington ratios larger than $10^{-2}$, 
we assume that the nuclear disc around the SMBH is thin and
radiative cooling is efficient. We  therefore assume that the luminosity of the disc can be described by the standard 
Shakura-Sunyaev disc model (\citealt{shak73}). Such sources will be highly luminous in X-rays.
We will refer to SMBHs in this regime as \textit{Shakura-Sunyaev discs} (SSDs).  For $\Ledd$ in the range
$ 10^{-4}-10^{-2}$, we assume that the nuclear accretion disc is thick and radiatively inefficient. We will refer to these SMBH as \textit{Advection Dominated Accretion Flows} (ADAFs) \citep{rees1982,narayan_yi1994,abramowicz1995}. By default, we assume that these sources make negligible
contributions to the X-ray luminosity function. Such sources are, however, expected to dominate radio source
counts. Finally, we classify  those with  $\Ledd< 10^{-4}$ as \textit{ inactive} and assume that such sources are
essentially undetectable against the emission of the host galaxy. 

Note that the choice of a threshold in the Eddington ratio to define both active states  
SSDs and ADAFs does not have a significant effect on the X-ray AGN luminosity functions as shown in Appendix~\ref{appendix:choiceaccretionregime}.

\subsection{Predicting X-ray observables}

\label{sub:method}

In this section, we describe the method used to predict X-ray luminosities from the SMBH accretion rates. 
We consider only AGN in the SSD regime ($\Ledd>10^{-2}$).  For such sources, the bolometric luminosity 
is
\be
 L_{\rm bol} = \frac{\epsr}{1-\epsr} \mdotbh c^{2}= \epsr \dot{M}_{\rm accr} c^2, 
\ee 
where $\epsr$ is the radiative efficiency and is set to $0.1$ as suggested by \cite{shak73}.

We use the redshift independent bolometric corrections of
\cite{marconi2004} to convert the  bolometric luminosity into intrinsic hard (2--10 keV) and soft (0.5-2 keV) X-ray band luminosities. The 
bolometric corrections are third degree polynomial relations defined as follows: 
\bary 
\log_{10} \Big( \frac{ \LhX }{\Lbol } \Big) & =& -1.54-0.24 L- 0.012 L^2 + 0.0015 L^3 \nonumber\\ 
\log_{10} \Big( \frac{ \LsX }{ \Lbol } \Big) & =& -1.64-0.22 L- 0.012 L^2 + 0.0015 L^3, 
\label{eq:bolcorrection} 
\eary 
where $L=\log_{10}(\Lbol/L_\odot)-12$. The bolometric corrections are computed
with a template spectrum that is truncated at $\lambda >1 \mu m$ to
exclude the IR bump \citep{marconi2004} produced by reprocessed UV
radiation. The correction is 
assumed to be independent of redshift. We note that \cite{vasudevan2009} have suggested
that the bolometric corrections may be a function of the Eddington ratio, but the
differences are not significant except in AGN with $\Ledd<10^{-2}$.
\cite{lusso2012} suggested that the bolometric corrections could be lower
than those of \cite{marconi2004} at high bolometric luminosities, but the offset is small
in the context of this work. \cite{hopkins2007} also proposed  expressions for the
bolometric corrections based on dust absorbed luminosities. For us, this is inappropriate since we base our analysis on
intrinsic X-ray luminosities. Thus, we opt for the relations of \cite{marconi2004} which  has 
the benefit of being consistent with previous studies. Hereafter, we always refer to luminosity $\LhX$ ($\LsX$) as the intrinsic luminosity
in the 2--10 kev (0.5--2 keV) rest-frame energy range. 
   
The emission of AGN may be absorbed if the circumnuclear environment
is rich in gas and dust. The absorption is likely to be highly anisotropic, making a fraction of sources 
undetectable in the soft X-ray band. From the observational data, it is unclear whether the obscured fraction
is a function of redshift and other sample properties. For example, early studies \citetext{e.g. \citealt{ueda2003}; 
\citealt{steffen2003}} did not find clear evidence for a redshift dependence, but recent
studies  have  established that the obscured AGN fraction increases with
increasing redshift \citetext{e.g. \citealt{hasinger2008};\citealt{treister2009};\citealt{ueda2014}}.  Because of these uncertainties, we prefer to compare the simulations
to observed soft X-ray luminosity functions for which the obscured fraction
has already been taken into account by simultaneously fitting to both hard and soft 
X-ray data \citep{aird2015}.  In future work, we will investigate the obscuration of AGN due to gas and dust,
taking into account the properties of the host galaxy.

\section{Results}
\label{sec:results}

\subsection{Properties of nearby SMBHs} 

\label{sub:bhmf_z0}
\begin{figure} 
\includegraphics[width = \columnwidth]{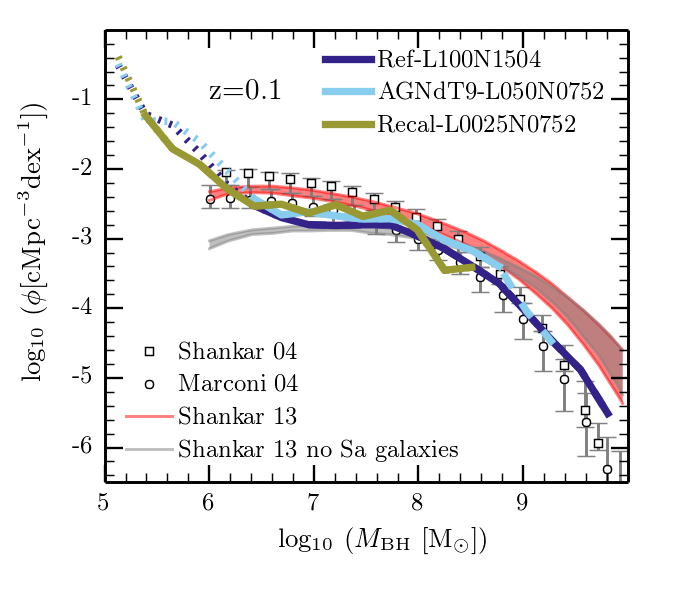} 
\caption{The SMBH mass functions of  \AGNdT~(light-blue), \REF~(blue) and \RECAL~(green) at $z=0.1$ The dotted part of each curve corresponds to SMBH masses below to the initial mass of
a gas particle. The dashed part at the high-mass end indicates the SMBH mass bins containing fewer than 10 objects per mass bin.  The 
grey region corresponds to the observational estimate of \protect\cite{shankar2013} who uses 
the $\mbh-\sigma$ relation from \protect\cite{mcConnell_ma2013}, while the red region corresponds to an estimate
 in which the SMBHs in the centre of Sa type-galaxies are included.  Older observational estimates from
\protect\cite{marconi2004} and \protect\cite{shankar2004} are shown as data points. The observational estimates 
all infer the black hole mass function indirectly and the differences are primarily driven by the choice 
of the $\mbh$-$\sigma$ calibration.}
\label{fig:bhmassfunction} 
\end{figure}

The SMBH mass function provides a useful overview of the SMBH population at low redshift.
To determine the average SMBH mass function with reduced sampling noise, we combine snipshot outputs as explained 
in \S~\ref{subsec:outputs}.  Fig.~\ref{fig:bhmassfunction} shows the SMBH mass function for 
the \REF~(dark blue line), \AGNdT~(light blue line) and \RECAL~(green line) simulations. In order to facilitate 
comparison with later plots and observational data, we include only the central black hole of each galaxy.
The level of agreement between simulations is good (better than 0.2 dex) for SMBHs with 
mass $>10^6 \msun$. This level of agreement is encouraging, but note that  
the EAGLE simulations were calibrated to reproduce the normalisation of
$\mbh$-$\mstar$ relation at $z=0$ \citepalias[see][Fig.~10]{schaye2015}. It is interesting to note the similarity of the high-mass end of the \REF~and \AGNdT~simulations.
Given the more effective AGN feedback in the \AGNdT~model, we might have expected a divergence at the 
massive end due to its greater gas mass loss from galaxy groups.
Clearly this is not the case. In Appendix~\ref{appendix:convergencetests}, we show that for SMBHs with mass $\lsim 1.6 \times 10^6 \msun$ 
the mass function depends strongly on the seed black hole mass and we indicate this region by dotted lines. 

It is interesting to compare the simulation SMBH mass functions to the estimates based on observational
data. Because SMBH masses can be directly determined only in an incomplete sample of galaxies \citep{kelly2013}, it is important to note that  the observational estimates of the SMBH mass function are indirect
and must be inferred from the correlations between SMBHs and the properties of
their host galaxy bulge. In the figure \ref{fig:bhmassfunction}, we show estimate from \cite{marconi2004}  (grey circles), \cite{shankar2004} (grey
squares) and more recent data from \cite{shankar2013} (red and grey regions). \cite{shankar2013} and
\cite{shankar2004} use the $\mbh$-$\sigma$ correlation, while \cite{marconi2004}
use the relation between SMBH mass and bulge
luminosity.  The simulated SMBH mass function is in reasonable agreement with the
observational estimates from \cite{shankar2004} and \cite{marconi2004} over a
wide mass range, but underestimates the abundance of the high-mass
SMBHs when compared to \cite{shankar2013}.  This discrepancy is somewhat surprising 
since both the simulation and \cite{shankar2013}  are calibrated to SMBH masses
from \cite{mcConnell_ma2013}. The discrepancy, and the variance between observational 
estimates, illustrates the uncertainty in deriving the SMBH mass function from 
observational data. Indeed, \cite*{shankar2013} note that
adopting different variations of the SMBH scaling relations leads to large
variations in the inferred SMBH mass function, since the scatter and mass range covered by the 
data must be taken into account. For example, \cite{shankar2013} show
that the low-mass end of their mass function depends strongly on how
the  $\mbh$-$\sigma$ correlation is applied to galaxies of different morphological
types. The red region assumes that the relation can be applied
regardless of morphology, while the grey region assumes that Sa
and late-type galaxies do not  have a SMBH. Thus, although there are
some differences between the simulated SMBH mass function and the more recent
observational estimates, these depend heavily on how the observational
calibration data is extrapolated. For this reason, it is far better to validate the simulation by comparing to the black hole mass-stellar mass relation
directly and in \cite*{schaye2015} we show that the simulation reproduces the observational data within their uncertainties.

Integrating the mass function, we obtain the predicted black hole mass density at $z=0.1$.
In the \REF~simulation, we find it to be $2.6 \times 10^5 \msun \Mpc^{-3}$, closely matching the 
observational value estimated by \cite{yu_tremaine_2002} ($2.6\pm 0.4\times10^5\msun \Mpc^{-3}$, adjusted to Planck Cosmology parameters), whose calculations are
based on the velocity dispersion of early-types galaxies in the Sloan Digital
Survey (SDSS).  This value is also consistent with the result from  \cite{aird2010} who used hard X-ray 
luminosities to compute the total energy released by the SMBH population through time ($2.2\pm 0.2\times10^5\msun\Mpc^{-3}$), although is 
lower than the estimate of \cite*{marconi2004} ($4.6^{+1.9}_{-1.4} \times 10^5 \msun\Mpc^{-3}$).   

\begin{figure} 
\includegraphics[width = \columnwidth]{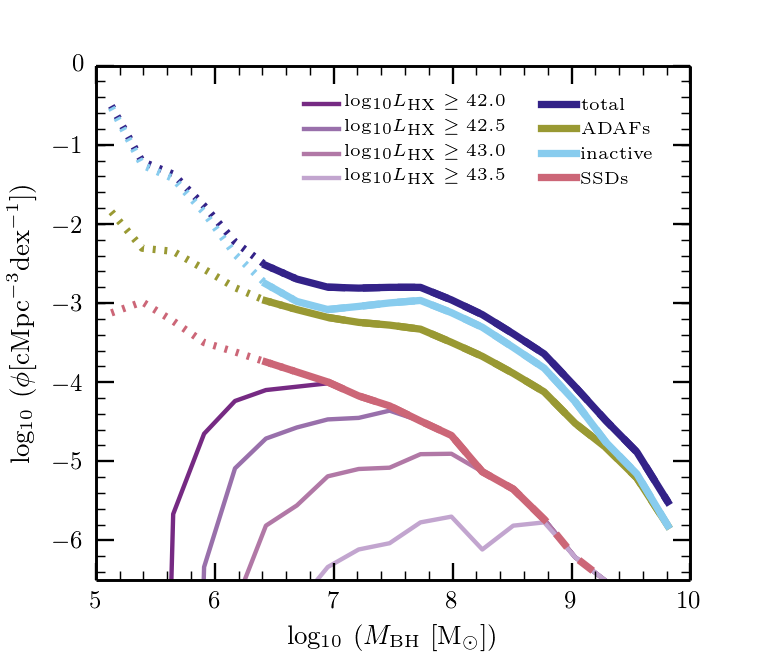} \\ 
\caption{The contributions of different
accretion regimes to the predicted SMBH mass function for the \REF~simulation at $z=0.1$.  
The dotted part of each curve 
corresponds to masses smaller than the initial mass of a gas particle and the dashed part of 
the curve to mass bins containing fewer than 10 objects. 
The dark blue line is the total SMBH population, the light blue line corresponds to inactive SMBHs ($\Ledd < 10^{-4}$), 
the green line to  SMBHs accreting as ADAFs ($10^{-4} \leq \Ledd <0.01$) and the red line to
SSDs ($\Ledd\geq 0.01 $). 
Purple lines in the
figure show the effect of also requiring the SSD sources to exceed a luminosity limit, as would be
the case in an observational survey.
The figure shows that inactive SMBH adominate the SMBH
mass function over a wide mass ranges, with a negligible contribution to the most massive SMBHs ($>10^{8}\msun$)
from SSDs. Comparing the contributions of different accreting SMBHs to the total SMBH mass function, 
the average `duty cycle' of SMBHs is determinated. The  predicted duty cycle for SSDs is $\sim
0.01$ in agreement with observational estimates. } 
\label{fig:bhmassfunction_accr} 
\end{figure}

In Fig.~\ref{fig:bhmassfunction_accr} we dissect the SMBH mass function according to accretion regime. We distinguish inactive SMBHs (light
blue), ADAFs (green) and SSDs (red). The three SMBH populations differ greatly in
normalization.  Most SMBHs are inactive, corresponding to $70\%$ of the total SMBH
population with $\mbh>10^7\msun$, while ADAFs (green line) correspond to $29\%$
and  SSDs correspond to only $\sim 1\%$ of the population.
The figure shows results from the \REF~simulation, but the breakdown is similar in the other simulations. 
Since SMBHs frequently switch between states between output times, we can view the 
differences in normalization as an average \textit{duty cycle}, and interpret the relative normalisation of the 
mass functions as the probability of finding a SMBH in any given state. 

The duty cycle, that is the fraction of the time a SMBH is active, shortens with increasing SMBH mass, with the probability of classifying
a SMBH as an active SSD varying from 0.05 for $\mbh\sim10^{6}\msun$ to  0.01 $\mbh\sim10^{8}\msun$. 
At higher masses, the probability of finding a present-day SMBH in the SSD state becomes extremely small.
Restricting the comparison to the SMBH population  with $\mbh>10^7\msun$, the SSD fraction is $0.02$ on average. This is consistent 
with the observational estimates of the average AGN lifetime for $\mbh<10^8\msun$ that corresponds to $3-13\times 10^7$ years
( e.g. \citealt{marconi2004}, \citealt{yu_tremaine_2002}). 

These trends are not particularly sensitive to the choice of the threshold used to define SSD systems; however, an observational survey will only detect black holes that exceed an X-ray luminosity (or flux) limit. Purples lines in Fig~\ref{fig:bhmassfunction_accr}  show the effect this has on the fraction of black holes that are detectable in an ideal hard (2--10~keV) X-ray survey. Our estimates account for bolometric corrections (as described in section~\ref{sub:method}) but do not account for additional selection effects, such as the difficulty of distinguishing faint AGN from emission associated with star formation. We focus on the results for an observational survey with a luminosity limit of  $\LhX>10^{43}\ergs$. For the highest mass black holes, all the SSD systems are detected, but below a black hole mass of $10^8\msun$, the observable population become increasingly biased. For black holes of mass $10^7\msun$, the detected population accounts for only 6\% of the SSD population and 0.3\% of all black holes of that mass.  As black hole masses drop below $10^6\msun$, the population becomes undetectable because of the Eddington limit. Fortunately, in practice, observational surveys are flux limited so that a range of luminosity limits can be probed; nevertheless, this exercise highlights the difficulty of constructing a complete census of the black hole population. We examine the X-ray emission of the simulation's black hole population in more detail in section~\ref{sub:HXRLF}.

\subsection{Properties of high-redshift  SMBHs }
Having examined the properties of SMBHs at low redshift, we now investigate the evolution of SMBHs.
We will  look at the evolution of the SMBH mass function, the
$\Ledd$-$\mbh$ distribution and the SMBH mass-halo mass relation. 
\label{sub:highz-bh}

\subsubsection{ The SMBH mass function}

\begin{figure} 
\includegraphics[width=\columnwidth]{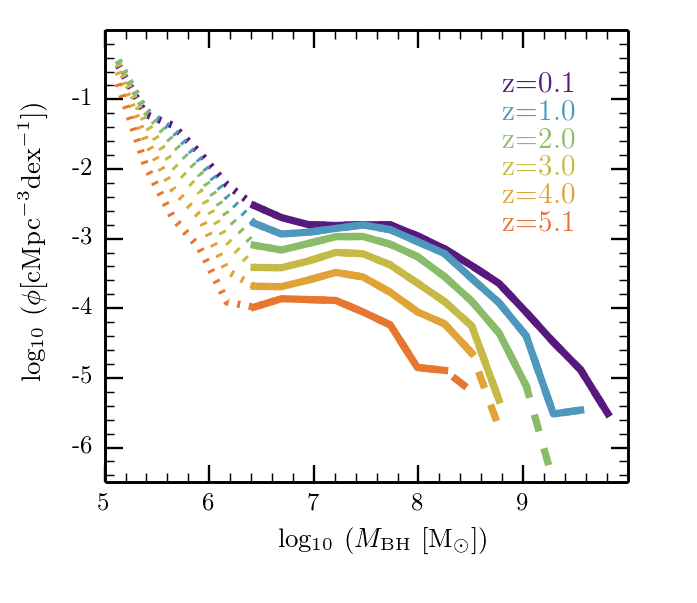} 
\caption{The evolution of the SMBH mass function from $z=5.1$ to $z=0.1$ 
in the \REF~simulation. The dotted part of the SMBH mass function corresponds to masses 
smaller than the initial mass of gas particles.
The dashed part corresponds to mass bins containing fewer than 10 objects. 
Colours represent different redshifts as indicated in the legend.  
The SMBH mass function shows a rapid evolution in  the normalization over the whole mass range from $z=5.1$ to $z=1$. Toward
lower redshifts ($z<1$),  the evolution is mostly restricted to a flattening of the slope at the high-mass end.}
\label{fig:bhmassfunction-evolution}
\end{figure}

\begin{figure*} 
\includegraphics[width=2\columnwidth]{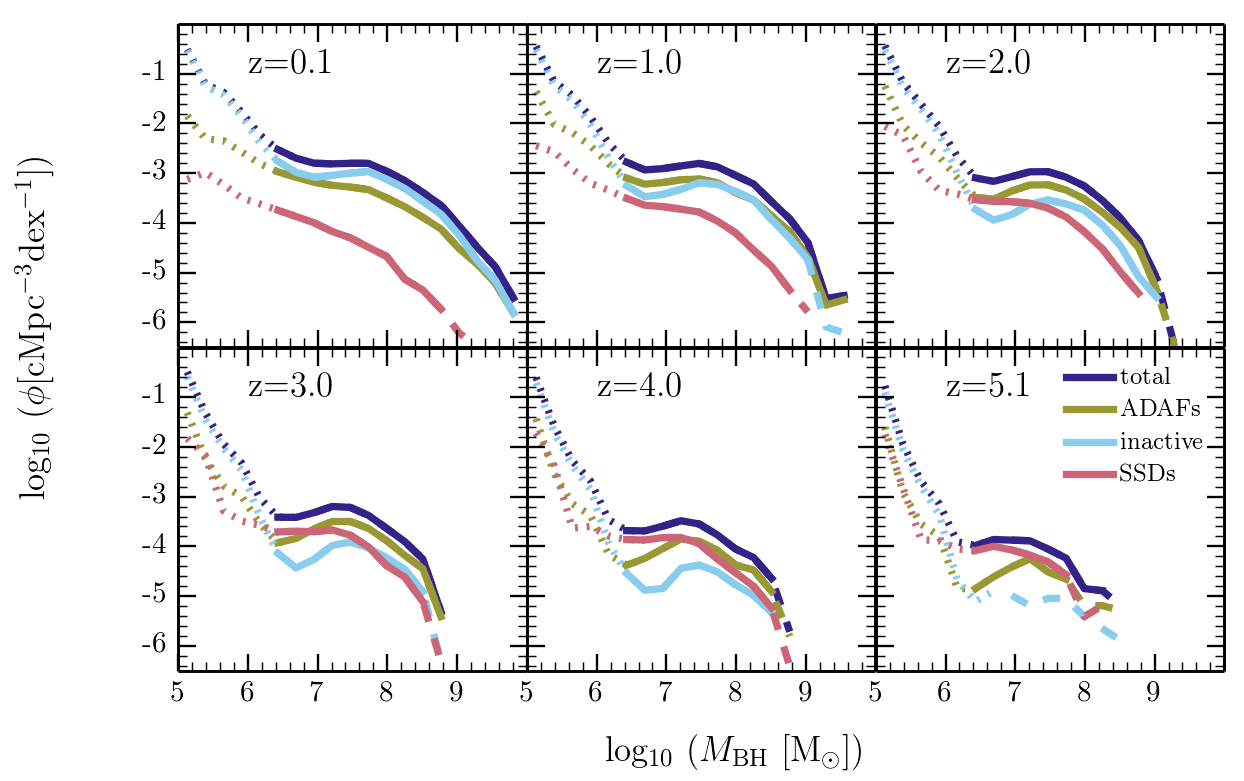} 
\caption{Evolution of the SMBH mass function split by accretion regime: light-blue lines correspond to inactive SMBHs ($\Ledd < 10^{-4}$),  green lines to  SMBHs accreting as ADAFs ($10^{-4} \leq \Ledd <10^{-2}$) and  red lines to
SSDs ($\Ledd\geq 10^{-2}$) from $z=0.1$ to $z=5.1$. At low redshift, the mass function is dominated by the inactive 
and ADAFs population, whereas at high redshift, the SSDs and ADAFs are the most abundant.}
\label{fig:bhmassfunction-evolution2}
\vspace*{\floatsep}

\includegraphics[width=2\columnwidth]{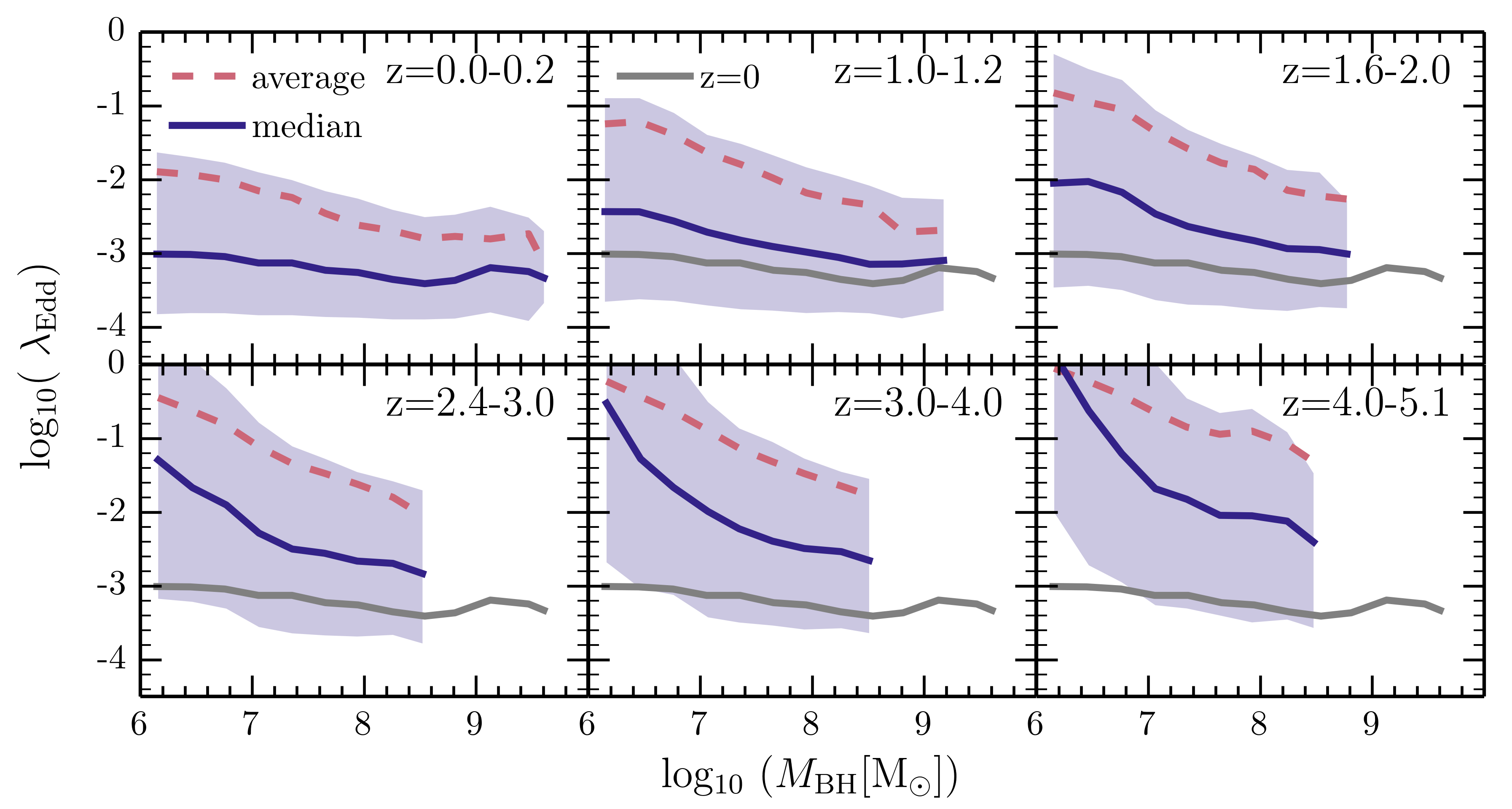} 
\caption{ Evolution of the Eddington ratio distribution ($\Ledd=\mdotbh/\mdotedd$), plotted as a function of 
black hole mass. Median and average $\Ledd$ are shown as solid blue and  dashed pink 
lines respectively. Only SMBHs with $\mbh>10^6\msun$ are shown.
The grey solid line repeats the median of the distribution at $z=0$. The coloured
region represents the 10$^{\rm th}$ and 90$^{\rm th}$ percentiles of the
distribution. $\Ledd$ increases as redshift increases, particularly for 
lower-mass black holes. SMBHs with $\mbh \lsim 10^7\msun$ have an increasing tendency
to be limited by the Eddington accretion rate at $z>2$.}
\label{fig:Ledddistribution} \end{figure*}


Fig.~\ref{fig:bhmassfunction-evolution} investigates the evolution of the SMBH mass
function. Between $z=5.1$ and $z=1.0$, the normalization of the SMBH mass function rapidly evolves by an 
order of magnitude. While the overall normalisation changes little at
lower redshifts ($z<1$), the abundance of the most massive objects increases as the break in the mass function becomes shallower with 
cosmic time, and the dip seen at low masses at intermediate redshifts is filled in.
The evolution of the SMBH mass function is similar to the evolution of the galaxy stellar mass function 
\citep[see][Fig.~2]{furlong2015a}.

In Fig.~\ref{fig:bhmassfunction-evolution2}, we show the evolution of the different accretion regimes. 
The contributions of ADAFs (green lines), SSDs (red lines) 
and inactive SMBHs (light blue) evolve relative to each other.  At $z<2$ the abundance of 
the SMBH mass function is dominated by ADAFs and inactive SMBHs, preserving a similar shape to
the total SMBH mass function. In contrast, the SSD population 
evolves rapidly in normalization from $z=2$ to $z=0.1$, also showing a rapid decrease in 
characteristic mass. This is a feature of AGN `downsizing' which we will return in the following sections. 
At $z>2$ the inactive SMBH population declines and the 
dominant populations are SSDs and ADAFs. The results described above show that there is switch in the 
nature of black hole accretion with cosmic time. Below $z=2$, the population is dominated 
by inactive SMBHs or ADAFs and only a tiny fraction is undergoing strong accretion. 
At high redshift, SMBHs undergo much more frequent high Eddington-rate accretion events.

\subsubsection{The $\Ledd$-$\mbh$ plane}

In Fig.~\ref{fig:Ledddistribution}  we show Eddington ratio, $\Ledd=\mdotbh/\mdotedd$, 
as a function of black hole mass in the  \REF~simulation from redshifts 0 to 5.  
Overall, the median Eddington ratio
decreases as a function of the black hole mass, with the SMBHs with mass $10^6-10^{7}\msun$ 
the most active population in the simulation through cosmic time.
However,  there is large scatter ($\sim 2$ dex) in the distribution of Eddington ratios for any
given SMBH mass due to the high variability of the mass
accretion rates. 
 
For a given SMBH mass, the median value of $\Ledd$ moves towards lower values as
redshift decreases. This trend is more evident in SMBHs with $\mbh<10^7\msun$,
where the median of log$_{10} \Ledd$ declines from $\sim-1$ at $z=3$  to
$-3$ at $z=0$. For SMBHs with higher mass, the median changes less dramatically, consistent with the difference in 
the evolution of the active and inactive SMBH populations shown in Fig.~\ref{fig:bhmassfunction-evolution2}.    
The figure also highlights that SMBHs of mass $< 10^7\msun$ have an increasing tendency to
be limited by the Eddington accretion rate at higher redshift, making it possible
to build quasar-mass black holes early in the history of the Universe.  In general, however, the 
SMBHs in the simulation accrete well below their Eddington limit.

\subsubsection{The $\mbh$-$\mcrit$ relation}
\label{sub:evol_bhm_hm_relation}
\begin{figure*}
\includegraphics[width=2\columnwidth]{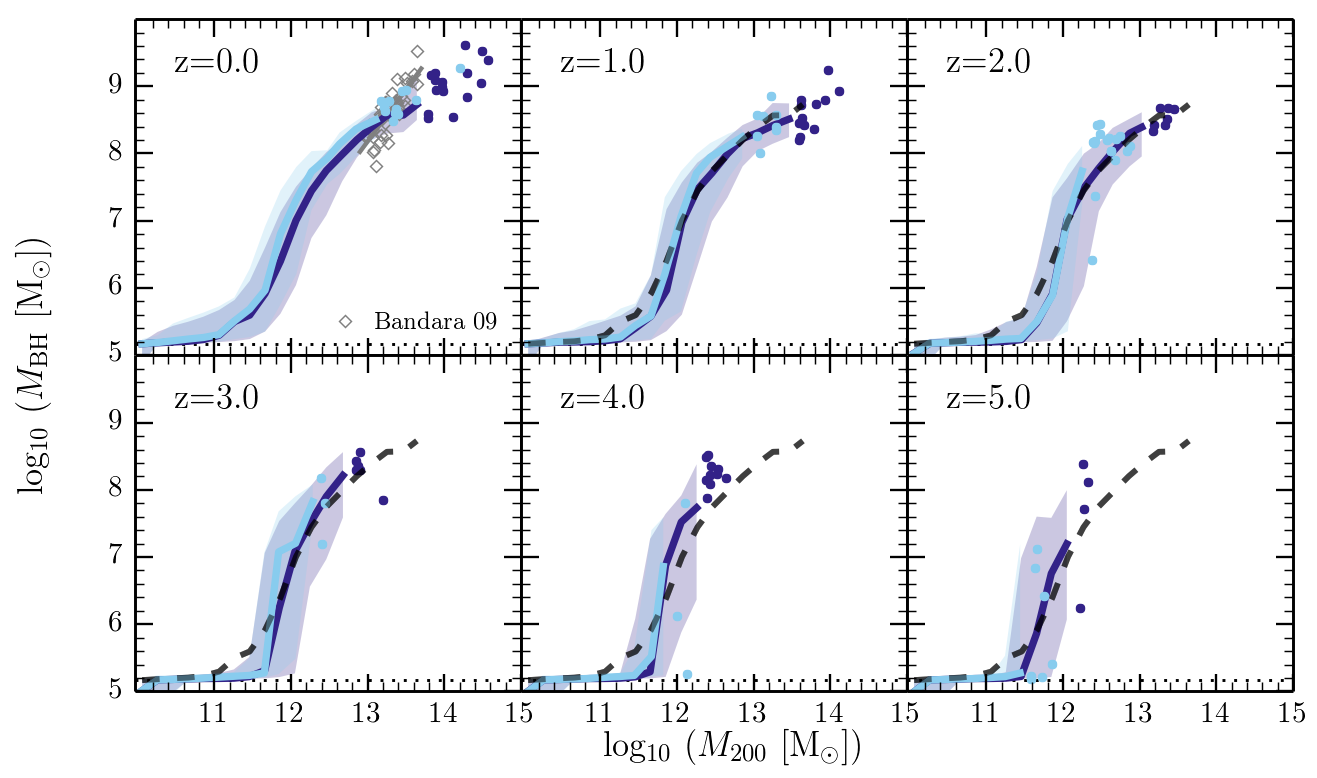}
\caption{The $\mbh$-$\mcrit$ relation for various redshifts from $z=0$ to $z=5$.
Solid lines represent the median for the simulations \REF~(dark
blue) and  \AGNdT~(light blue). Circles show individual haloes for 
bins with less than 10 haloes per dex mass bin. Coloured regions represent the 10$^{\rm th}$
to 90$^{\rm th}$ percentiles of the distribution, and the dark dashed line represents
the $\mbh$-$\mcrit$  relation at $z=0$. Grey diamonds show observational
estimates from \protect\cite{bandara2009} and the grey solid line the fit to the $\mbh$-$\mcrit$  relation found 
in that paper. The dotted horizontal line shows the BH seed mass. The $\mbh$-$\mcrit$  relation undergoes a transition
for halos with $ \mcrit \sim 10^{11.6} \msun$. The transition evolves little with redshift, showing only
a tendency to be slightly more abrupt at higher redshifts.}
\label{fig:bhmass_halomass_evol}
\end{figure*}

Fig.~\ref{fig:bhmass_halomass_evol} shows the evolution of the
the central SMBH mass -- halo mass relation for different redshifts from $z=5$ to $z=0$.
   
We also include the  estimates  at $z=0$ from
observations of strong gravitational lenses by \cite{bandara2009} as grey diamonds and the grey line. In
comparison to the simulations, \cite{bandara2009} find a steeper relation, but this calculation
was based on the assumed form of the $\mbh$-$\sigma$ relation  and not on direct
observations of the mass of the central SMBHs. 

The two simulations agree closely, and both show rapid BH growth in the
halo mass range $10^{11.5}$-$10^{12.5}\msun$ demonstrating that this rapid growth phase does not depend on the 
details of the SMBH feedback scheme (or indeed the SMBH seed mass, as we show in Appendix~\ref{appendix:convergencetests}).  The origin of this transition will be investigated in Bower et al.( 2016, in preparation), showing that it emerges as a result of a change
in the hot gas content of the halo (see also \citealt{bower2006}). In halo masses below the transition, the gas content of galaxies
is regulated by stellar feedback; however, in more massive haloes, supernova-driven outflows stall as the hot halo becomes established and the gas content of the galaxy is regulated by the more energetic SMBH driven feedback.
The median $\mbh$-$\mcrit$ relation evolves very little with redshift, so that
at  $z=2$ many massive SMBHs have already been assembled and  a
population of halos with $\mcrit>10^{12}\msun$ already host SMBHs with
$\mbh>10^{8.5}\msun$. At higher redshifts the transition between the SMBH mass
regimes becomes more abrupt, and SMBHs in this regime must grow rapidly as their halo mass increases
which is consistent with the increasing median Eddington ratio seen in Fig.~\ref{fig:Ledddistribution}. 
SMBHs in higher-mass haloes ($>10^{12}\msun$) have released enough energy into the host halo to ensure that 
the cooling time becomes long, and the galaxy is starved of further fuel for star formation. This later process results in a self-regulated
growth as noted by \cite*{booth_schaye2010} leading, together with the BH growth due to mergers, to the small scattering well-defined slope seen in Fig.~\ref{fig:bhmass_halomass_evol}.

\subsection{Observable diagnostics of SMBH growth}
\label{sub:AGNlf}
In this section we investigate observables related to  gas accretion onto
SMBHs. We will focus on the
hard and soft X-ray AGN luminosity functions and the evolution of the
space density of AGN in hard X-rays through cosmic time.  

\subsubsection{Hard X-ray luminosity functions} 
\label{sub:HXRLF}

\begin{figure*} 
\includegraphics[width=2\columnwidth]{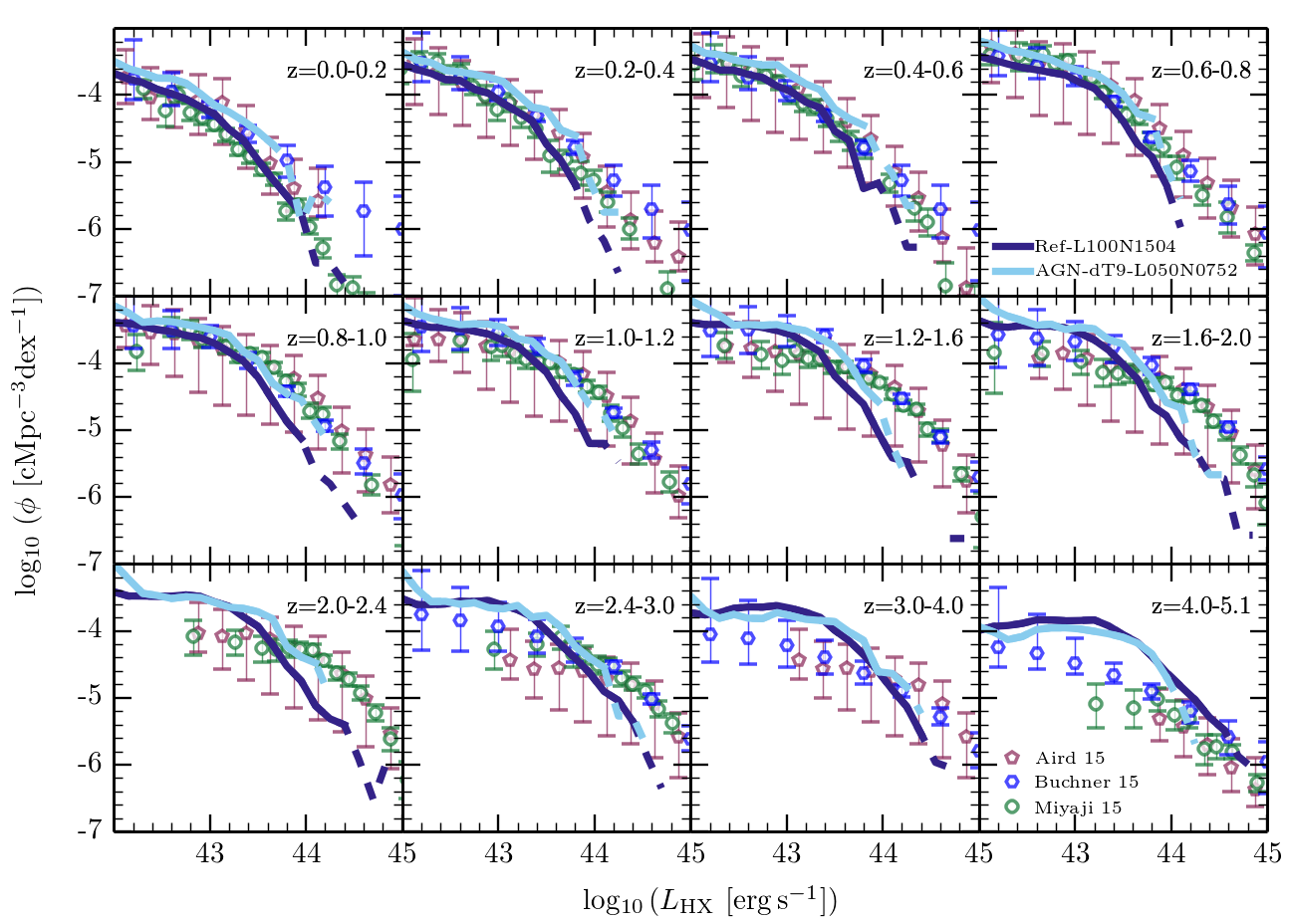}
\caption{Evolution of the hard (2--10~keV)  X-ray luminosity functions in simulations \REF~(dark blue) and \AGNdT~(light blue). Each panel corresponds to a different redshift bin as indicated. The simulation curves are dashed where there are fewer than 10 AGN per dex luminosity bin.   Green circles correspond to the observational
estimates from \protect\cite{miyaji2015}, red pentagons to
\protect\cite{aird2015} and bright blue hexagons  to \protect\cite{buchner2015}. 
Comparing the simulations to each other, we find good agreement at all redshifts with
differences no larger than 0.2 dex in normalisation. The abundance of AGN of a given luminosity increases up to $z\approx2$ 
and then declines. Compared to the observations,  the
simulations match the data well for $z<0.8$, but for 
$1.2<z<4.0$ they underestimate the abundance of AGN with luminosities greater than
$\LhX>10^{44} \ergs$, and may overestimate the abundance of fainter sources.  The differences for brighter sources are, however, affected by the limited volume of our simulation.} 
\label{fig:HXRLF}
\end{figure*}

\begin{figure*} 
\includegraphics[width=2\columnwidth]{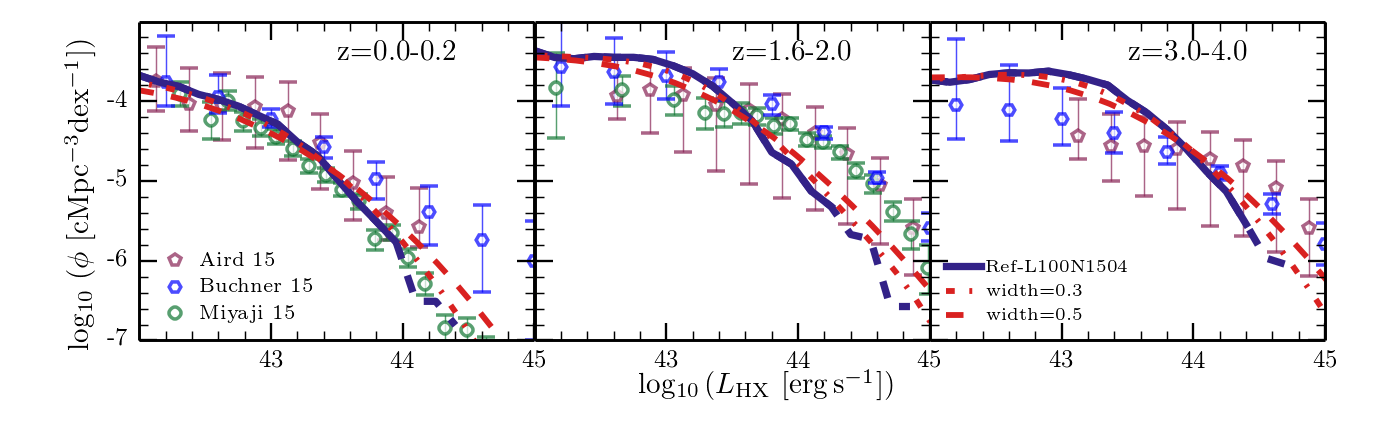}
\caption{An illustration of the plausible impact of unresolved AGN variability on the hard X-ray 
luminosity function in EAGLE. Blue solid lines and points reproduce respectively model \REF~and observations of Fig.~\ref{fig:HXRLF}. 
Red dash-dotted  and dashed lines represent the hard X-ray luminosity
function convolved with a Log-normal distribution of width $0.3$ and $0.5$ dex
respectively to illustrate the impact of variability not resolved by
the simulations. The mean of convolution kernel is set in order to conserve the total energy released.
Unresolved variability does not significantly alter the comparison between the observed luminosity 
function and the simulation prediction.} 
\label{fig:HXRLF_2}
\end{figure*}

Fig.~\ref{fig:HXRLF} shows the predicted hard (2--10~keV) X-ray luminosity function
(HXRLF) over the redshift range $0$ to $5$. Intrinsic X-ray luminosities have been derived using the bolometric corrections 
described in Sec.~\ref{sub:method}, and include only SMBHs in the SSD regime. We show two simulations,
\REF~(dark blue lines) and \AGNdT~(light blue lines). 
The HXRLF shows strong evolution in shape and normalization in both simulations.
Below $z=2$ the simulations agree with each other  within $0.2$ dex in normalisation, with \AGNdT~slightly above the HXRLF of \REF. At $z>2$ the simulations are still similar but present higher discrepancies in the bright end of the HXRLF. The bright end of the HXRLF may be affected  the limited statistics available in our volume or by variability on short timescales that is not resolved. 
 
We also compare the predicted HXRLF to recent observational estimates
based on the deep X-ray fields from \cite{miyaji2015} (green circles),
\cite{aird2015} (red pentagons) and \cite{buchner2015} (bright blue hexagons).  Overall,  the comparison between the observations and simulations is encouraging, given the simplicity of the subgrid model used. We stress that the  parameters of SMBH growth and AGN feedback were calibrated to match the stellar mass function at $z=0.1$ and the normalization of 
the $\mbh$-$\mstar$ relation, not to match the evolution of AGN.  
Looking in detail, the observations are in good agreement (within the observational error bars) 
out to $z=0.8$; however,  at $1.2<z<4.0$ and higher luminosities ($\LhX>10^{44}\ergs$), the simulation HXRLF appears to decline
 with $\LhX$ more quickly than seen in the observations. The discrepancy does not appear to be due to the sampling statistics but we cannot entirely rule out the possibility that it is due to the 
finite volume of the simulation (see Appendix~\ref{appendix:convergencetests}) because we sample a relatively small number of massive galaxies in the simulations.  Above $z\sim2$, there is an overabundance of low luminosity  ($\LhX<10^{43}\ergs$) AGN in EAGLE. In this regime,  however, the observational constraints are quite uncertain, as can be seen by comparing different observational datasets.

We have mentioned previously that one possible factor that affects the HXRLF is
the variability of AGN that is not resolved in the simulation. Short time-scale variations could originate
from fluctuations in accretion disk viscosity, for example. In galactic binary systems such as stellar remnant compact objects, 
order of magnitude variations in the accretion rate arise from the ionisation instability in accretion disks, and similar instabilities may be present in SMBHs \citep{done2007}. Such `flickering' cannot be resolved in our simulations which only attempt
to model variations in  the gas supply rate on $10^2$ pc scales. 

An illustration of the effect of short-timescale variability is shown in Fig.~\ref{fig:HXRLF_2}. Here, we show the effect of convolving source luminosities with a log-normal distribution with $\sigma$ between 0.3 and 0.5 dex per luminosity. 

We have chosen the log-normal distribution as a simple way to illustrate the potential impact of flickering since we are interested in the effect of order of magnitude variations in source luminosity. It is important to note that the central value of the convolution kernel has been set so that the average (expectation) luminosity is independent of $\sigma$. We have explored relatively high $\sigma$ values in order to assess the maximum impact of unresolved variability in the simulation. A source with $\sigma=0.5$ is, instantaneously, an order of magnitude brighter or fainter than the mean luminosity for 5\% of the time, and a factor of 3 brighter or fainter for 32\% of the time. Values higher than 0.5 would imply that the instantaneous $\LhX$ is almost unrelated to the SMBH accretion rate.

Solid lines reproduce the \REF~simulation and observational data from Fig.~\ref{fig:HXRLF}. The effect of including unresolved variability is  shown as red dot-dashed and red dashed lines in Fig.~\ref{fig:HXRLF_2}. As the width of the convolving Gaussian is increased intrinsically low luminosity sources scatter in the high luminosity bins, and the simulation tends to agree better with the observational data. The overall effect is relatively small, however, and does not seem able to reconcile the simulation with the observational data.
The convolution has little effect on the fainter luminosities ($\LhX<10^{43}\ergs$) and so cannot account for the overabundance of faint sources in the simulation at $z>4$. We have already stressed that the observational measurements are uncertain in this regime.

\begin{figure} 
\includegraphics[width=1\columnwidth]{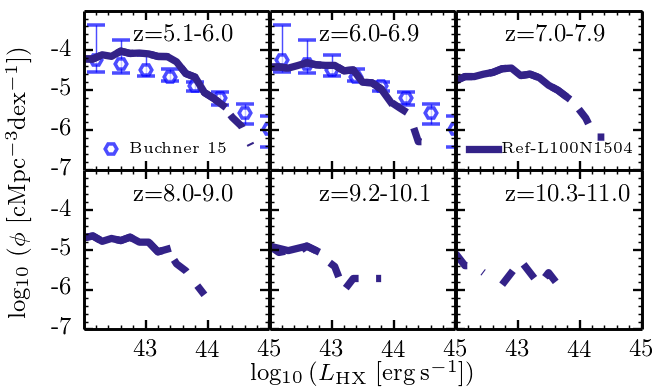} 
\caption{Hard X-ray luminosity functions of AGNs from $z=5.1$ to $z=11$. The HXRLF amplitude  and shape  evolve 
rapidly as redshift decreases. We compare the simulations with observations from \protect\citealt{buchner2015} (open points with error bars), finding good agreement at $z=5.1$ and $7$ over the luminosity range that can be compared.  No observational data is available at higher redshift.}
\label{fig:highHXRLF} 
\end{figure}

Although the volume of the simulation is rather small for the characterisation of extreme high redshift events, 
for completeness we show the predicted hard X-ray luminosity function in EAGLE at $z=5-11$ in Fig.~\ref{fig:highHXRLF}.
We see that the HXRLF amplitude decreases with redshift and evolves rapidly in shape. Above $z=8$, the simulation suffers 
from particularly poor sampling for AGNs with $\LhX>10^{43}\ergs$ (indicated by the dashed line). 
Comparing to observational estimates from \cite{buchner2015}, we find reasonable agreement between $z=5.1$ and $7$
over the luminosity range that we are able to probe. Observational data is not available at higher redshifts in this luminosity 
range, and the figure presents the model predictions.

\subsubsection{Soft X-ray luminosity function}
\label{sub:SXRLF}

Soft (0.5--2~keV)  X-ray measurements provide a useful complement to hard X-ray surveys. 
The evolution of the soft X-ray AGN  luminosity function (SXRLF) has been
investigated by e.g. \cite{miyaji2000,hasinger2005,aird2015}. A complication, however, is that an
uncertain fraction of sources will be obscured by the gas column along the line of sight through the host galaxy to 
the AGN. Rather than trying to model the effects of obscuration in the simulation, we compare to the 
results of \cite*{aird2015} in which the effects of obscuration have been empirically corrected.  

\begin{figure*} \includegraphics[width=2\columnwidth]{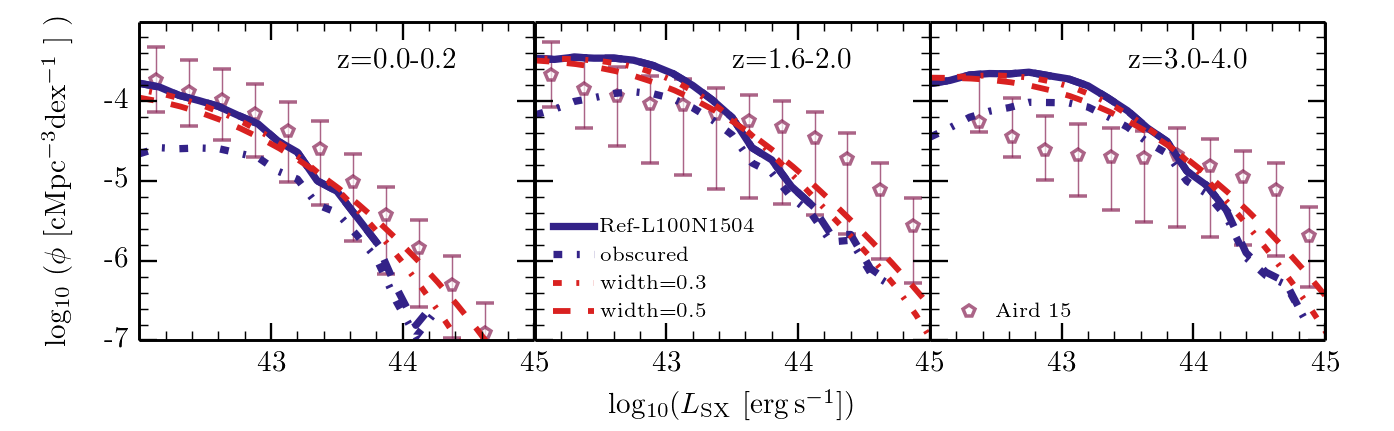}
\caption{Evolution of the soft X-ray luminosity function. Each panel represents the SXRLF at
different redshifts from $z=0-4$ as indicated. Solid lines represent the SXRLF
in the simulation \REF, with blue dashed lines showing the luminosity bins containing fewer than 10 
AGN per dex luminosity bin. Blue dot-dashed lines represent the SXRLF when empirical obscuration estimates
are applied to the simulation data. The simulation results are compared to 
the observational estimate of \protect\cite*{aird2015} (red pentagons) which includes a correction for obscuration
and should thus be compared to the solid lines from the simulation. 
Red lines represent the SXRLF convolved with Log-normal luminosity variations of $0.3$ (dot-dashed lines) and $0.5$ (dashed lines ) in dex, 
bringing  the simulation into slightly better agreement with the observations. Discrepancies in the abundance at fainter luminosities 
could be the result of a more rapid evolution of the obscuration, 
however, similar discrepancies are seen in Fig.~\ref{fig:HXRLF}.
} \label{fig:SXRLF} \end{figure*}

Fig.~\ref{fig:SXRLF} shows the SXRLF for the simulation \REF~(solid lines).  
To illustrate the importance of obscuration, we have included blue dot-dashed  
lines to show the expected abundance of detectable (i.e. unobscured) objects using the prescription of 
\cite{hasinger2008}. Comparing the SXRLF with obscuration to the one without, the fraction of
obscured AGN can vary between $0.83$ and $0.01$ with the largest values 
found at low luminosities ($\LsX<10^{42}\ergs$) and the smallest at high luminosities ($\LsX>10^{44}\ergs$). Note, however, 
that these ratios are not a theoretical prediction of the simulations, but the effect of corrections derived from 

We compare the predicted SXRLF  (solid lines) to the observational estimates
from  \cite*{aird2015} (red pentagons). The observed counts are corrected for the effects of obscuration 
by comparing the hard- and soft-band X-ray data. The data points should therefore be compared
to the solid lines derived from the simulations. The simulation broadly reproduces the observed evolution 
across cosmic time, particularly for the faintest part of the SXRLF
($\LsX< 10^{42.5-44}\ergs$) at $z<2$. Even at low redshift, the brightest part of the SXRLF is steeper than
observed, but the discrepancy is only greater than the observational uncertainties in the region where we have fewer than 10 sources per bin (shown as dashed lines). 

As we discussed above, we would expect some additional contribution from unresolved variability, and we show the effect of log-normal luminosity variations between width of 0.3 and 0.5 dex as red dot-dashed and red dashed  lines respectively. As we found in Fig.~\ref{fig:HXRLF_2}, this has relatively little impact.
Towards high redshifts ($3<z<4$), the simulations predict a higher amplitude of the SXRLF (particularly at luminosities 
of $\LsX\sim 10^{43}\ergs$). This might indicate an overabundance of the faint AGN in the simulations, but may also be due to a greater redshift dependence of obscuration than accounted for by \cite{aird2015}.

However, a similar over-abundance is seen in  both soft and hard X-ray luminosity functions, and in general  in both X-ray bands the luminosity functions follow a similar evolution. It seems therefore that the offset between the simulation and the observations must either be real (in the sense that the numerical implementation of SMBH accretion used in the simulations generates an excess of low luminosity sources) or be due to observational selection effects (for example, we have not attempted to model observational selection effects such as the difficulty of detecting faint
AGN against the galaxy's nuclear star formation).

In general terms, the evolution of the SXRLF in the
simulation  evolves in broad agreement with the observational constraints, opening a window to explore more deeply
the connection between  BH accretion rates, obscuration and the gas and star formation properties of galaxies.

\begin{figure} 
\includegraphics[width=1.1\columnwidth]{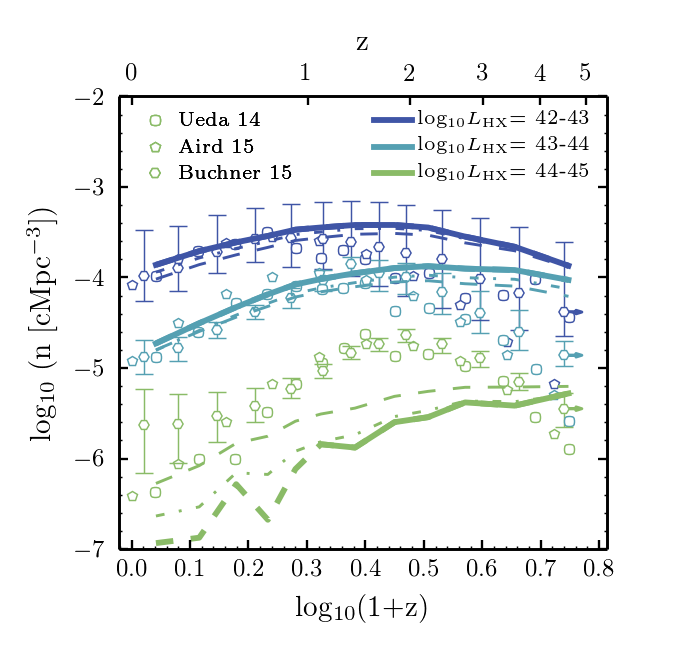}
\caption{Evolution of the comoving  number density of AGN in
simulation \REF,  broken into three different hard X-ray luminosity bins: blue solid lines
correspond to  AGN with $\LhX=10^{42-43}\ergs$, light blue lines to AGN with
$\LhX=10^{43-44}\ergs$ and green lines to $\LhX=10^{44-45}\ergs$. The solid lines become dashed  when there are less than 10 objects per 
dex luminosity bins. Observational
estimates from \protect\cite*{ueda2014}, from \protect\cite*{aird2015} and the values obtained by 
integrating the observational estimates from \protect\cite{buchner2015} . The hexagons with arrows represent 
values that where estimated within a smaller redshift bin.  The 
evolution of the comoving number density of AGN with $\LhX=10^{42-43}\ergs$
and $\LhX=10^{43-44}\ergs$ is similar to the observed
estimates for $z\lsim 1.5$ but declines more slowly with redshift than suggested by the observations at $z>2$. The abundance of the brightest AGN 
is affected by the size of the simulation and additional variability not captured by the simulation. Thinner lines
 illustrate the effect of convolving AGN luminosities with log-normal flickering of 0.3 and 0.5 dex.}
\label{fig:evolution-HXLF} 
\end{figure}

\subsubsection{Evolution of the comoving number density of AGN}
\label{sub:downsizing}

Fig.~\ref{fig:evolution-HXLF} shows the evolution of the comoving number density of AGN in the simulation \REF~(solid lines). As we discussed above, some AGN variability may not be accounted for in the
simulation and we illustrate the effect of convolving source luminosities with a log-normal distribution of width 0.3 to 0.5 dex. We split the  AGN population into three luminosity bands as indicated in the legend.  This figure reproduces the information already seen in Fig.~\ref{fig:HXRLF_2}, 
but allows us to investigate the evidence for AGN `downsizing' more clearly. In the observational data, higher luminosity sources peak in abundance at 
progressively higher redshifts. A similar trend is
seen in the simulations: brighter AGN ($\LhX=10^{44}-10^{45}\ergs$;
green lines) peak at redshifts greater than 3 while fainter AGN ($\LhX=10^{42}-10^{43}\ergs$; dark blue lines) peak at $z\approx 1.4$.  

For comparison, we show recent estimates from \cite*{ueda2014} and \cite*{aird2015}. We also show values obtained  by integrating the luminosity functions from \cite*{buchner2015}. There is reasonable agreement between
simulations and observations for the lower luminosity bins, $\LhX=10^{42}-10^{43}\ergs$ (dark blue line)
and $\LhX=10^{43}-10^{44}\ergs$ (light blue line), out to $z\sim1.5$. Moving towards higher
redshifts, the simulations  predict too  many faint AGN in comparison to
observations, the discrepancy becoming almost 0.8 dex in the comoving number density at
$z\sim 5$. We note, however, similar
discrepancies are seen in observational data due to uncertainties in detecting faint AGN at high redshifts.
For example, the comoving number density of AGN of $\LhX=10^{42}-10^{43}\ergs$ from \cite{ueda2014} (blue
circles, $10^{-4.4}\,\cMpc^{-3}$) is higher by 0.8 dex compared to \cite{aird2015} (blue pentagons, $10^{-5.2}\,\cMpc^{-3}$)  at $z\sim5.0$. 
Moreover, \cite*{giallongo2015} recently reported a more abundant population of faint  AGNs at $z=4$-$6$  by studying  AGN candidates from the multiwavelength CANDELS deep surveys \citep{grogin2011,koekemoer2011}. The sample was initially selected by the UV rest frame of the parent galaxy and thus is able to account for sources with marginal X-ray nuclear detections. This population might extend to even higher redshift ($z>10$). \cite*{madau2015}, based on this result, suggest that such  contribution of active galaxies drives the reionization of hydrogen and helium satisfying several observational constrains such as the observed flatness of  HI photoionization rate between $z=2$ and $z=5$ and the estimated integrated Thompson scattering optical depth ($\tau=0.056$) found in the Lymann $\alpha$ opacity of the intergalactic medium and cosmic microwave (CMB) polarization.

The comoving number density of the brightest AGN is low in the simulations compared to the 
observational estimates. However, the comoving number density of the brightest AGN can be  affected 
by additional AGN variability combined with the low numbers of bright AGN in the finite simulation volume.
As we show in the figure, convolution with log-normal flickering of 0.5~dex goes some way to account for the high abundance of 
bright AGN seen in the observations.

Overall, while the `downsizing' trend is present in the simulations, it is not as clear as suggested by the observational data. In particular, the abundance of AGN of a particular luminosity has a broader plateau than suggested by observations, principally because the rapid decline observed in the abundance of faint AGN at high redshift
 seen in observations is shallower in the simulations. At almost all redshifts, however, the simulations tend to underpredict the abundance of the brightest AGN, and their peak of their abundance occurs at too high redshift.  We should be careful not to over-interpret this apparent discrepancy, however, since the abundance of these objects is poorly sampled in the simulation volume.

\section{Conclusions}
\label{sec:discussions}

We have examined the evolution of supermassive black holes (SMBHs) across cosmic time predicted by
the EAGLE simulations \citepalias{schaye2015,crain2015}.  The EAGLE project consists of a suite of
hydrodynamical simulations with state-of-the-art subgrid models of galaxy
formation including radiative cooling, star formation, reionization, abundance evolution, stellar evolution and mass loss, feedback from star formation,  and
SMBH growth and AGN feedback. The parameters of these subgrid models were calibrated to reproduce the observed galaxy mass function
and sizes at $z=0.1$. In particular, the efficiency of AGN accretion and feedback were set to reproduce the break in the stellar mass function at $z=0.1$ and the normalization of the SMBH mass-stellar mass relation at $z=0$.  It is important to emphasise
that the subgrid models of SMBH growth and AGN feedback  do not make any explicit distinction between {\it quasar} and {\it radio} modes, and that we only distinguish sources with high and low Eddington ratios during the analysis. The main findings are summarised as follows:

\begin{itemize}

\item  The main properties of  nearby SMBHs are reproduced well in the EAGLE simulations. Within the observational uncertainties, the $z=0$ SMBH mass function is similar to estimates from \cite*{shankar2004}, \cite*{marconi2004} and \cite*{shankar2013}, and the density of SMBHs in the local Universe is also comparable to that observed. This agreement is partly the result of the calibration strategy (see Fig.~\ref{fig:bhmassfunction}). As a post-processing step, we divide the present-day black hole mass function as a function  of Eddington ratio, $\Ledd$. We associate sources with $\Ledd \geq 10^{-2}$ with X-ray luminous \textit{Shakura-Sunyaev discs} (SSDs), sources with $10^{-4} \leq \Ledd <10^{-2}$ as \textit{ Advection Dominated Accretion Flows} (ADAFs) and classify that sources with $\Ledd < 10^{-4}$ as inactive. At low redshift the mass function is dominated by inactive and ADAF black holes (Fig.~\ref{fig:bhmassfunction_accr}). Assuming that SMBHs cycle between the SSD, ADAF and inactive states, we estimate that the {\it duty cycle} for SMBHs in the SSD state is $\sim 0.01$, which is comparable to the observational estimates (e.g. \citealt{soltan1982}, \citealt{marconi2004}, \citealt{yu_tremaine_2002}).      

\item The mass function of SMBHs in the EAGLE simulation evolves rapidly in amplitude from $z>5$ to $z=2$. At redshift $2$ a large fraction of the  $\mbh >10^7\msun$ population has already been formed (Fig.~\ref{fig:bhmassfunction-evolution}). Between $z=2$ and the present-day  the mass function evolves more gradually in normalisation. When we break this evolution down by accretion state, we find that luminous SSD systems, while a minor contribution at the present-day, become increasingly dominant at high redshift (Fig.~\ref{fig:bhmassfunction-evolution2}). This trend can also be clearly seen by examining the evolution of the Eddington ratio distribution directly (Fig.~\ref{fig:Ledddistribution}).

\item We examine the dependence of black hole mass on the dark matter halo mass, $\mcrit$ (Fig.~\ref{fig:bhmass_halomass_evol}). The $\mbh$--$\mcrit$ relation has a characteristic shape, with SMBH masses growing little above the seed mass in haloes less massive than $\sim10^{12}\msun$, but showing a sharp rise in more massive haloes. The fast growth of the SMBH ends when its mass exceeds $\mbh \sim10^{8}\msun$. SMBHs follow an almost linear trend with $\mcrit$ in larger haloes. The characteristic shape of this relation evolves little with redshift, with a suggestion that the steep rise in mass becomes more abrupt as redshift increases. 
   
\item The black hole mass function, the Eddington ratio distribution and the SMBH dependence on halo mass cannot be directly observed and must be inferred by combining observational surveys with, for example, a calibration between black hole mass and stellar mass. To facilitate a more direct comparison between the model and observational data, we compute the X-ray luminosity function in the rest frame for SMBHs in the SSD state. We use bolometric corrections from \cite*{marconi2004} to convert the bolometric output predicted by the model
into the intrinsic hard and soft X-ray luminosities. We compare the hard-band X-ray luminosity functions with the observational measurements from \cite*{miyaji2015}, \cite*{aird2015}  and \cite*{buchner2015} (Fig.~\ref{fig:HXRLF}). The finite volume of the simulation limits the comparison to hard X-ray AGN luminosities lower than $\LhX\sim 10^{44}\ergs$. At low redshifts, the 
simulations agree extremely well with the observational data. At higher redshift ($z>1$) the simulations tend to underpredict the abundance of high luminosity sources with $\LhX\gsim10^{44}\ergs$, although we cannot rule out the possibility that this is a result of the simulation's limited volume. At $z=2$ and above, the amplitude of the predicted luminosity function  
appears higher than observed, particularly around $\LhX\sim10^{43}\ergs$, although some caution is required since we have not attempted to include observational 
selection effects (other than bolometric corrections) in our predictions.  We find a similar result when we compare to obscuration-corrected soft-band X-ray luminosity functions
 from \cite*{aird2015}.

\item  The hard X-ray luminosity functions we derive include the effect of variability captured by the model due to gas flows on kpc scales, but unresolved variability (for example due to flickering in the accretion disk) that may cause additional fluctuations in luminosity. Given the steepness of the X-ray luminosity function, this could have an important impact.  However, we present a simple model based on additional log-normal distributed flickering to show that this has only a limited impact on the comparison with observational data (Fig.~\ref{fig:HXRLF_2}).

\item We investigate AGN {\it downsizing}  in the simulation (Fig.~\ref{fig:evolution-HXLF}). The observed trend seen in observational data is qualitatively reproduced: the comoving  number density of higher-luminosity AGN peaks at higher redshift, and the simulations 
are in good quantitative agreement with the observational data for $\LhX<10^{44}\ergs$ 
at $z\lsim 2$. At higher redshifts, the simulations produce more active SMBHs than observed, resulting in
 a shallower roll-over of the AGN abundance. The finite volume of the simulations and the possible effects of flickering make difficult to reliably compare the abundance of the more luminous AGN. Taken at face value, 
the simulations do not predict the rapid rise in the abundance of the brightest ($\LhX>10^{44}\ergs$) objects seen in some observational surveys between $z=0$ and $z=2$. Larger volume simulations, and a better understood model for AGN flickering, are required to determine if this is
due to  a real discrepancy between the hydrodynamical model and the observational data.
\end{itemize}

The results we find are broadly consistent with other simulations and semi-analytic calculations. For example, 
\cite*{fanidakis12} used a version of the semi-analytic GALFORM code, similar to that of \cite{bower2006}, in which galaxy formation is approximated
as a network of analytic differential equations that are applied to halos that grow in a dark matter N-body simulation. It is assumed that
SMBHs grow either by accretion from the diffuse gas halo, if this is stable against cooling, or as a result of  gas flows produced during merger and disk instability driven 
starbursts. The semi-analytic model is able to probe large volumes and hence more luminous sources, and the model indeed
generates a population of very luminous sources at high redshift, improving the match to the observational data. Similarly to EAGLE, theses calculations do not show the strong `down-sizing' trend inferred from the observational data and the authors conclude that the perceived evolution is largely the result of obscuration. 

Recently, \cite*{hirschmann2014} and \cite*{sijacki2015} have presented an analysis of black hole properties in large volume cosmological simulations. \cite*{hirschmann2014} combined simulations of a $500$ Mpc region at 
low resolution (a factor of $\sim 100$ higher  than the particle mass of the \REF~simulation) run to $z=0$, with the results 
obtained from a 68~Mpc region with a resolution similar to that of the EAGLE simulation, but run only to $z=1$. Their prescription of AGN feedback is extensively
based on \cite{springel2005a} and two explicit modes of AGN feedback are assumed, namely, a quasar and a radio mode with a switchover depending on the source Eddington ratio. The efficiency of feedback in  radio mode is $4$ times larger than  that in quasar mode. These models generally fit the 
observed AGN luminosity functions reasonably well, although they also find that the abundance of $\LhX\sim10^{43}\ergs$ sources tends to be 
overestimated when the resolution of their large volume calculation is increased.  Accounting for the differences in resolution and volume, their results appear compatible with our own.  

\cite*{sijacki2015} presented an analysis of SMBHs in the Illustris simulation \citep{vogelsberger2014} which, while similar in volume and resolution of our work, differs greatly in its implementation of AGN feedback and accretion onto SMBHs. In particular, the Illustris simulation employs different schemes for feedback in high and low Eddington ratio sources. In low accretion states radio feedback is implemented by depositing energy in thermal `bubbles' at  some distance from the central galaxy \citep{sijacki2007}. In this high accretion mode feedback energy is dumped at the location of the BH at every timestep, a procedure that is expected to result in significant radiative losses at this resolution (e.g. \citealt{dallavecchia_schaye12}) The black hole mass function derived from the Illustris simulations differs 
significantly in shape from that in EAGLE, being essentially a power-law that increases in steepness as a function of mass. At low redshift, rare SMBHs (more massive than $\mbh\sim10^9\msun$) are more abundant, but most  of these SMBHs accrete at rates less than $10^{-4}$ of the Eddington limit. The results for the luminosity functions are broadly similar to ours (within the uncertainty of the data that are shown). In particular, they also find that the model tends to overpredict the abundance of moderate luminosity ($\LhX\sim10^{43}\ergs$) AGN at $z>2$. In terms of AGN downsizing, the model does appear to capture the rapid decline of the abundance of high-luminosity sources, although it is unclear whether this is largely affected by the selection of sources based on an Eddington ratio criterion of $\Ledd > 10^{-4}$,
 given the significant difference in the black hole mass functions of the simulations. 

Although most of  our qualitative results seem compatible with these  earlier works. It is important to stress the greater
simplicity of the AGN feedback model used in EAGLE, and the fact that basic galaxy properties like stellar masses and sizes are better reproduced. It is therefore encouraging that the model, which uses a single  mode of AGN feedback and in which  AGN feedback energy is a fixed fraction of the accretion rate, 
captures so many of the trends seen in observational data. The results we have presented from the EAGLE simulations open a new window to investigate the co-evolution of the SMBH growth and galaxy evolution. In future work, we will investigate more consistently the obscuration of AGN due to gas and dust by
including the properties of the host galaxy. We will also investigate the effects of AGN feedback on the host galaxies and how this evolves through cosmic time.

\section*{Acknowledgements}

Authors thank Johannes Buchner, James Aird, Francesco Shankar and Takamitsu Miyaji  for  providing their observational data and the 
referee for the useful comments. This work would have not be possible without Lydia Heck's technical support. YRG thanks Johannes Buchner
for his useful comments about observations.  YRG gratefully 
acknowledges financial support from the Mexican Council for Science and Technology (CONACYT) (Studentship No. 213183) and partial support from Center of Excellence in Astrophysics and Associated Technologies (PFB 06).
This work used the DiRAC Data Centric system at Durham
University, operated by the Institute for Computational
Cosmology on behalf of the STFC DiRAC HPC Facility
(www.dirac.ac.uk). This equipment was funded by BIS National E-infrastructure capital grant ST/K00042X/1, STFC
capital grant ST/L00075X/1, and STFC DiRAC Operations grant ST/K003267/1 and Durham University. DiRAC
is part of the National E-Infrastructure. We also gratefully acknowledge PRACE 
for awarding us access to the resource Curie based in France atTres Grand Centre de
Calcul. This work was sponsored by the Dutch National
Computing Facilities Foundation (NCF) for the use of
supercomputer facilities, with Financial support from the
Netherlands Organization for Scientific Research (NWO).
The research was supported in part by the European Research Council under the European Union's Seventh Framework 
Programme (FP7/2007-2013) / ERC Grant agreements 278594-GasAroundGalaxies, GA 267291 Cosmiway,
and 321334 dustygal, the Interuniversity Attraction Poles
Programme initiated by the Belgian Science Policy OWNce
([AP P7/08 CHARM]), the National Science Foundation under Grant No. NSF PHY11-25915, the UK Science and Technology 
Facilities Council (grant numbers ST/F001166/1 and ST/I000976/1.
 We thank contributors to SciPy \footnote{http://www.scipy.org} , Matplotlib \footnote{http://www.matplotlib.sourceforge.net} , and the Python programming language \footnote{http://www.python.org}.




\appendix 

\section{Convergence Tests }

\label{appendix:convergencetests}
Following the discussion of the numerical convergence in
\citetalias{schaye2015}, we use this appendix to investigate the impact of  the simulation volume on AGN observables.
We also present the effects of varying the initial seed SMBH mass and simulation resolution on the AGN luminosity functions.  
The simulations that we consider are described in Table \ref{table:tests} and Table \ref{table:simulations}. 

\begin{figure*} \begin{tabular}{ccc}
\includegraphics[width=0.7\columnwidth]{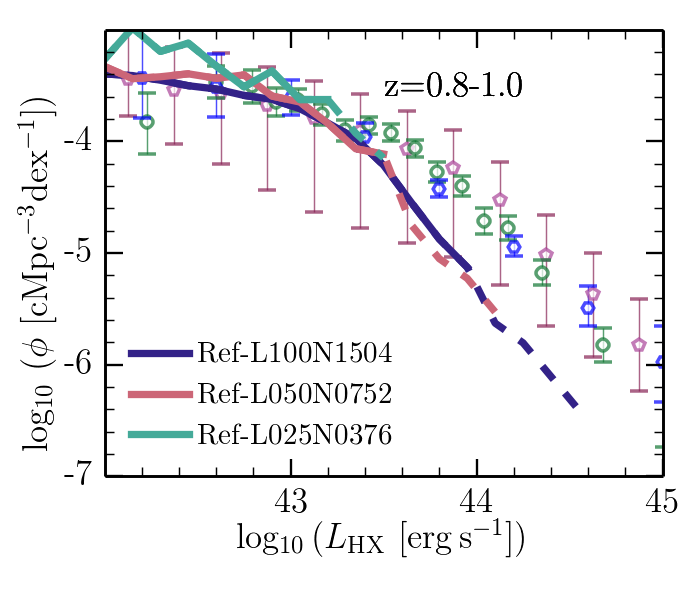} &
\includegraphics[width=0.7\columnwidth]{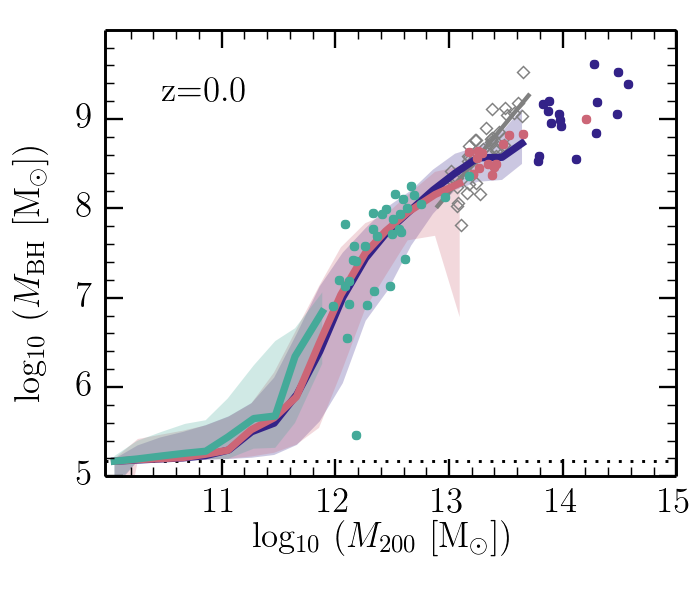} &
\includegraphics[width=0.7\columnwidth]{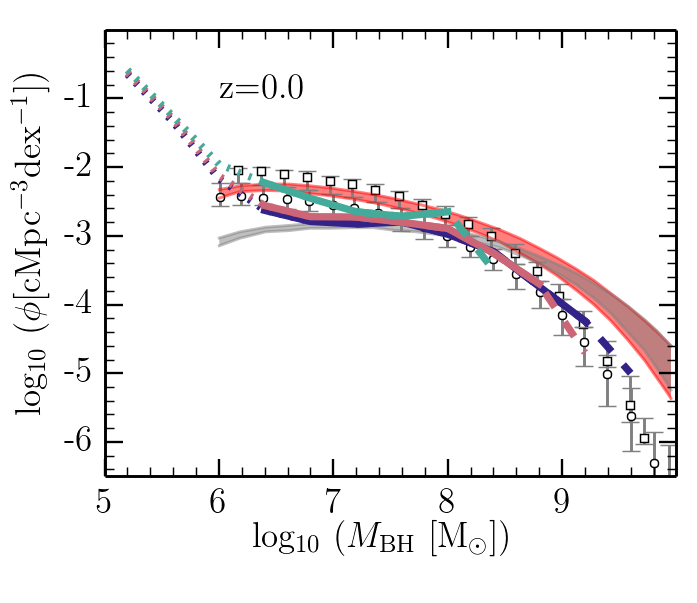} \\
\end{tabular} \caption{ Convergence by box size. {\it Left panel}: The HXRLF at $z=0.8-1.0$ in the
simulations \REF~(blue),  \REFFIFTY~(pink) and Ref-L025N0376 (green). The discrepancies between
simulations \REF~and \REFFIFTY~are smaller than $0.2$ dex for $\LhX<10^{44}\ergs$.{\it Middle panel}: the median of
the  $\mbh$-$\mcrit$  relation at $z=0$, with shaded regions showing
the 10$^{\rm th}$ and 90$^{\rm th}$ percentiles of the distribution. In each panel, observational data is presented
 following Figs.~1, 5 and 7. {\it Right panel}: the SMBH mass function at $z=0$. There is good consistency between
 results in different simulation volumes.}
\label{fig:boxsizetest} \end{figure*}

Fig.~\ref{fig:boxsizetest} investigates the sensitivity to the volume of the simulation of the hard X-ray luminosity function
(HXRLF) at $z=1$, the median of the  $z=0$ SMBH mass-halo mass relation  and the SMBH
mass function  at $z=0$. Dash lines show the limit of our sampling statistics. The left panel of 
Fig.~\ref{fig:boxsizetest} shows a good convergence in the HXRLF. The
discrepancies for the HXRLF are smaller than 0.2 dex between simulations in
the brightest part of the HXRLF. The plots in the figure also show excellent agreement within these limitations.

\begin{figure*} 
\begin{tabular}{ccc}
\includegraphics[width=0.7\columnwidth]{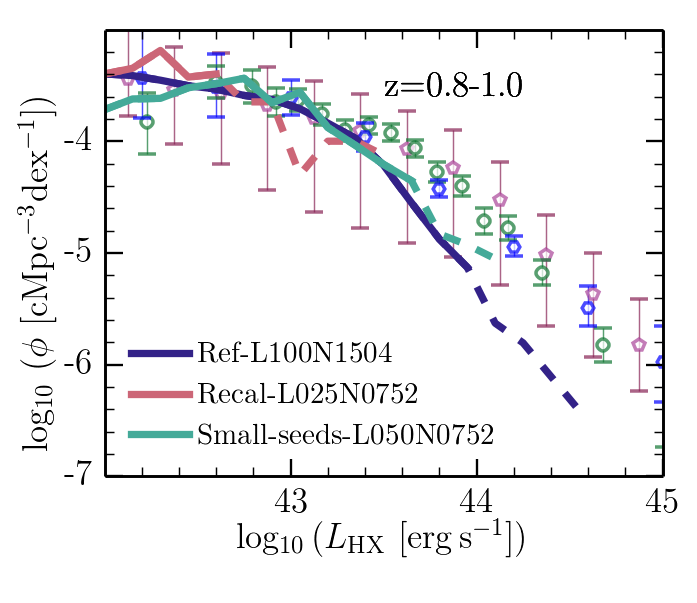} &
\includegraphics[width=0.7\columnwidth]{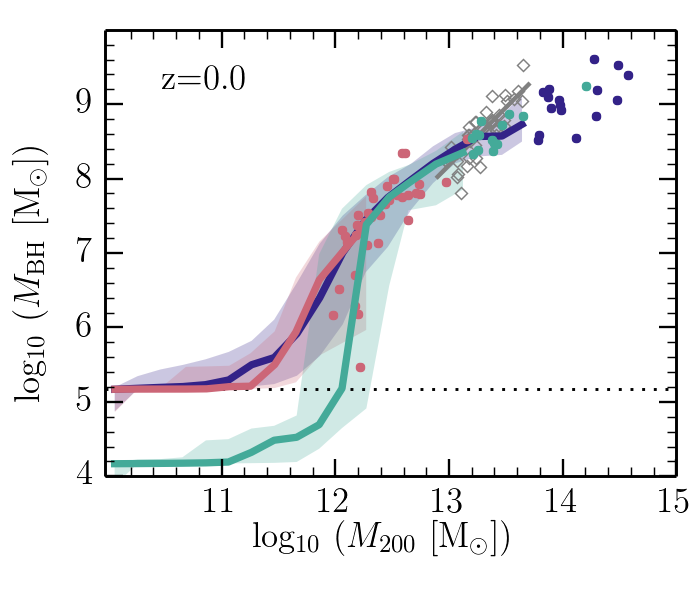} &
\includegraphics[width=0.7\columnwidth]{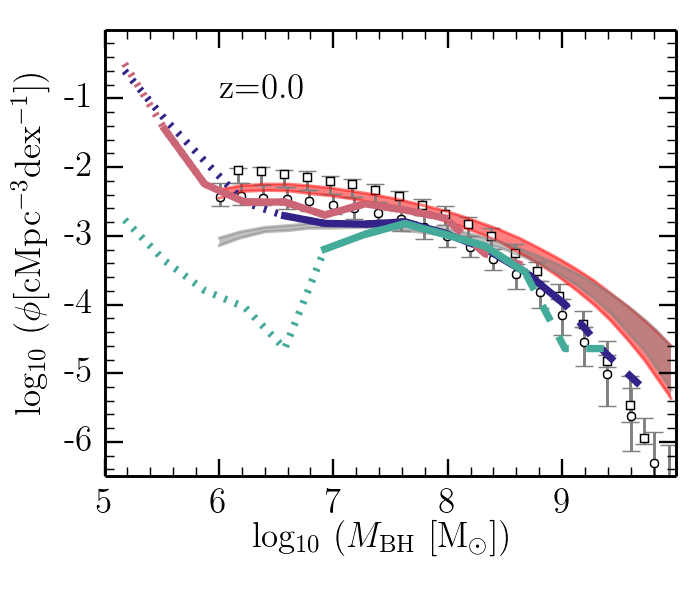}  \\

\end{tabular}
\caption{Similar to Fig.~\ref{fig:boxsizetest} but showing the simulations \REF~(blue), \RECAL~(pink) and \SmallSeeds~(green).  The panels investigate the dependence on resolution and  on the assumed SMBH seed mass. 
The $\mbh$-$\mcrit$  distribution 
presents a steep rise in halos with $\mcrit\sim10^{11.5-12.5}\msun$ in all simulations,
but is sharper in \SmallSeeds, while the SMBH mass function and HXRLF are largely unaffected.}
\label{fig:sbgridtest} \end{figure*}

Fig.~\ref{fig:sbgridtest} shows the same panels, but compares the 
simulations \REF~(blue), \SmallSeeds~(pink) and \RECAL~(green). Although we compare  different box sizes, 
the previous figures show that this is not a concern providing that the sampling statistics are appropriately accounted for.
The left panel shows that the HXRLF is insensitive to resolution and to a change in the model for AGN feedback. Decreasing 
the SMBH seed mass by an order of magnitude has only a small effect (compared to observational uncertainties) at the faintest luminosities shown.
The middle panel of Fig.~\ref{fig:sbgridtest} shows the median of the distribution of the $\mbh$-$\mcrit$  relation at $z=0$. The median  and scatter of each simulation shows a similar shape, however, the simulation \SmallSeeds~presents a sharper rise in halos of $10^{11.5-12.5} \msun$ in comparison to the other simulations.  This results from the SMBHs in \SmallSeeds~having to grow faster to reach the self-regulated 
$\mbh$-$\mcrit$  relation. Note however, that this steep rise persists in the three simulations, indicating that this fast growth of 
SMBHs in such haloes are independent of the subgrid parameters of the simulation. The right panel presents the SMBH mass function at $z=0$ in simulations \REF, \SmallSeeds~and \RECAL.  The agreement between \REF~and \RECAL~is 
better than 0.2 dex, comparable to the differences in the galaxy mass functions of these simulations. In contrast, the lower mass end of the SMBH mass function ( $\mbh<10^{7.5}\msun$) is strongly affected by the SMBH seed mass. The simulation \SmallSeeds~predicts lower values
for the SMBH mass function by $\sim 1$dex for SMBHs with mass smaller than $10^{7}\msun$. Nevertheless, the massive end of the SMBH mass function present an impressive level of agreement between simulations.  

\begin{table*}
\caption{Box length, initial particle number, initial baryonic and dark matter particle mass, comoving and maximum proper gravitational softening 
for the EAGLE simulations used in this paper. }
\begin{tabular}{|l|l|l|c|c|c|c|}
\hline
Name& L               &   $N$             &     $m_{\rm g}$         &  $m_{\rm DM}$ &  $\epsilon_{\rm com}$ & $\epsilon_{\rm prop}$ \\
    & [$\rm cMpc$]    &                   & [$\msun$]           & [$\msun$] &     $[\rm ckpc]$ &          $[\rm ckpc]$ \\ 
\hline
Ref-L100N1504 & $100$  & $2\times1504^3$  &  $1.81\times 10^6$& $9.70\times10^6$&  $2.66$     &  $0.70$ \\                

Ref-L050N0752 & $50$  & $2\times752^3$  &  $1.81\times 10^6$& $9.70\times10^6$&  $2.66$     &  $0.70$ \\                

\SmallSeeds  & $50$  & $2\times752^3$  &  $1.81\times 10^6$&  $9.70\times10^6$&  $2.66$     &  $0.70$ \\                

Ref-L025N0376  & $25$  & $2\times376^3$  &  $1.81\times 10^6$&  $9.70\times10^6$&  $2.66$     &  $0.70$ \\                

Recal-L025N0752 & $25$  & $2\times752^3$  &  $2.26\times 10^5$& $1.21\times10^6$&  $1.33$     &  $0.35$ \\               
\hline
\end{tabular}
\label{table:tests}
\end{table*}

\section{Choice of Accretion Regimes}
\label{appendix:choiceaccretionregime}
\begin{figure*} 
\includegraphics[width=2.0\columnwidth]{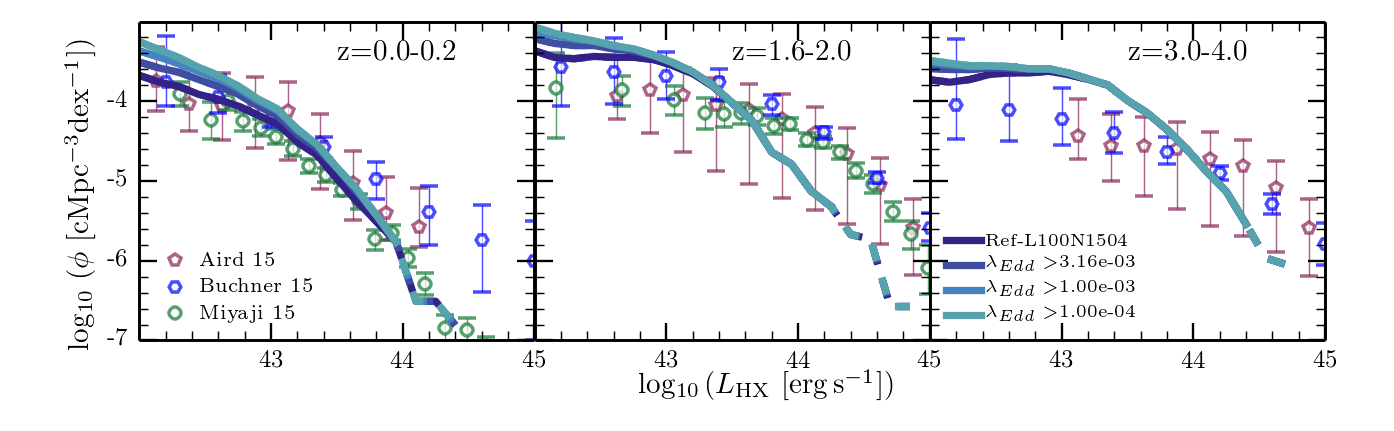}
\caption{Similar to Fig.~\ref{fig:HXRLF_2} but showing the simulation \REF~(blue) with increasing value of $\Ledd$ of SMBHs to be consired SSDs. Different colours indicate different value of $\Ledd$ as indicate the legend. SMBHs above of this limit are considered in the HXRLF. The HXRLF are largely unaffected by this limit, specially the bright end of the HXRLF.}
\label{fig:meddtest} \end{figure*}

In section \ref{sec:accretion-regimes}, we define two active 
accretion regimes that depend on the value of Eddington ratio. We assume that black holes  with $\Ledd$ that are higher than $10^{-2}$  are luminous sources of X-rays (since the nuclear disc is thin and radiative cooling efficient) and consider them as SSDs. Lower luminosity sources are assumed not to contribute to the X-ray luminosity functions we show in the main text. In this appendix, we explore the impact of varying this limit.
Fig.~\ref{fig:meddtest} shows the HXRLF considering SMBH to be X-ray luminous when $\Ledd$ is larger than a minimum that varies from $10^{-4}$ to $10^{-2}$. We show redshifts from $z=0$ to $z=4$. The dependence is weak, especially for the bright end of HXRLF.  For the faint end of the HXRLF, (AGN with $\LhX$ lower than $10^{43} \ergs$) the difference becomes of $\sim 0.5$ dex, comparing $\Ledd> 10^{-2}$ and $\Ledd>10^{-4}$. This discrepancy becomes smaller with increasing redshift and is smaller than the observational error bars.

\bsp	
\label{lastpage}

\begin{thebibliography}{99}
\bibitem [\protect\astroncite{Abramowicz et al.}{1995}]{abramowicz1995} 
{Abramowicz}, M.~A. , {Chen}, X. , {Taam}, R.~E., 1995, ApJ, 452, 379.

\bibitem [\protect\astroncite{Aird et al.}{2010}]{aird2010} 
{Aird}, J. , {Nandra}, K. , {Laird}, E.~S. and {Georgakakis}, A. , 
{Ashby}, M.~L.~N. , {Barmby}, P. , {Coil}, A.~L. , {Huang}, J.-S. , 
{Koekemoer}, A.~M. , {Steidel}, C.~C. and {Willmer}, C.~N.~A.
2010, MNRAS, 401, 2531.

\bibitem [\protect\astroncite{Aird et al.}{2015}]{aird2015} 
Aird, J., Coil, A.~L., Georgakakis, A., et al.\ 2015, arXiv:1503.01120

\bibitem [\protect\astroncite{Angles-Alcazar et al.}{2016}]{angles-alcazar2016}
Angl{\'e}s-Alc{\'a}zar, D., Dav{\'e}, R., Faucher-Gigu{\`e}re, C.-A., {\"O}zel, F., \& Hopkins, P.~F,, 2016, arXiv:1603.08007 

\bibitem[\protect\astroncite{Bah{\'e} et al.}{2016}]{bahe2016} 
Bah{\'e}, Y.~M., Crain, R.~A., Kauffmann, G., et al.\ 2016, \mnras, 456, 1115 
 
 
\bibitem[\protect\astroncite{Bandara et al.}{2009}]{bandara2009} 
Bandara,~K., Crampton,~D., Simard, ~L., 2009, ApJ, 704, 1135. 

\bibitem[\protect\astroncite{Binney \& Tabor}{1995}]{binney_tabor1995} 
Binney J., Tabor G., 1995, MNRAS, 276, 663. 

\bibitem[\protect\astroncite{Bondi \& Hoyle}{1944}]{bondi44} 
Bondi ~H., Hoyle ~F., 1944, MNRAS, 104, 273.

\bibitem[\protect\astroncite{Booth \& Schaye}{2009}]{booth_schaye2009} 
Booth C.~M., Schaye ~J., 2009, MNRAS, 398, 53.

\bibitem[\protect\astroncite{Booth \& Schaye}{2010}]{booth_schaye2010} 
Booth C.~M., Schaye ~J., 2010, MNRAS, 405, L1.




\bibitem[\protect\astroncite{Bower et al.}{2006}]{bower2006} Bower R.~G.,
Benson A.~J., Malbon R., Helly J.~C., Frenk C.~S., Baugh C.~M., Cole S., Lacey
C.~G., 2006, MNRAS, 370, 645. 



\bibitem[\protect\astroncite{Buchner et al.}{2015}]{buchner2015} 
Buchner J., Georgakakis A., Nandra K., et al. 2015, ApJ, 802, 89. 


\bibitem[\protect\astroncite{Chabrier}{2003}]{chabrier2003} 
Chabrier, G., 2003, PASP, 115, 763.

\bibitem[\protect\astroncite{Choi et al.}{2013}]{choi2013} 
Choi, E., Naab, T., Ostriker, J.~P., Johansson, P.~H., \& Moster, B.~P., 2014, MNRAS, 442, 440.

\bibitem[\protect\astroncite{Churazov et al.}{2001}]{churazov2001} 
Churazov E., Br{\"u}ggen M., Kaiser C.~R., B{\"o}hringer H., Forman W., 2001, ApJ, 554, 261. 



\bibitem[\protect\astroncite{Crain et al.}{2009}]{crain09}
Crain R.~A., Theuns ~T., Dalla Vecchia ~C., Eke V.~R. , Frenk S.~C., Jenkins ~A., Kay S.~T. and Peacock J.~A. {\it et al.}, 2009, MNRAS,  399, 1773.

\bibitem[\protect\astroncite{Crain et al.}{2015}]{crain2015}
Crain, R.~A., Schaye, J., Bower, R.~G., et al.\ 2015, arXiv:1501.01311 

\bibitem[\protect\astroncite{Croton et al.}{2006}]{croton2006}
Croton D. ~J., Springel V., White S. ~D. ~M., De Lucia ~G., Frenk S. C., Gao L., Jenkins ~A. and Kauffmann ~G. {\it et al.}, 2006, MNRAS, 365, 11.

\bibitem[\protect\astroncite{Cullen \& Dehnen}{2010}]{cullen_dehnen2010} 
Cullen, L. and Dehnen, W., 2010, MNRAS, 408, 669.  


\bibitem[\protect\astroncite{Dalla Vecchia \& Schaye}{2012}]{dallavecchia_schaye12} 
Dalla Vecchia  ~C. and Schaye ~J., 2012, MNRAS, 426, 140.




\bibitem[\protect\astroncite{De Lucia \& Blaizot}{2007}]{delucia2007} 
De Lucia G., Blaizot J., 2007, MNRAS, 375, 2


\bibitem[\protect\astroncite{Di Matteo et al.}{2008}]{dimatteo2008}
Di Matteo, T. and Colberg, J. and Springel, V. and Hernquist, L. and  Sijacki, D., 2008, ApJ, 676, 33.


\bibitem[\protect\astroncite{Done et al.}{2007}]{done2007} 
Done, C., Gierli{\'n}ski, M., \& Kubota, A.\ 2007, A\&ARv, 15, 1

\bibitem[\protect\astroncite{Durier \& Dalla Vecchia}{2012}]{durier_dallavecchia12} 
Durier, F. and Vecchia D.~C., 2012, MNRAS, 419, 465.



\bibitem[\protect\astroncite{Fanidakis et al.}{2012}]{fanidakis12}
 {Fanidakis} N. and {Baugh} C.~M. and {Benson} A.~J. and {Bower} R.~G. and 
{Cole} S. and {Done} C. and {Frenk} C.~S. and {Hickox} R.~C. and 
{Lacey} C. and {Del P.~Lagos} C., 2012, MNRAS, 419, 2797.


 
\bibitem[\protect\astroncite{Ferland et al.}{2013}]{ferland2013} 
G. J. Ferland, R. L. Porter, P. A. M. van Hoof, R. J. R. Williams, N. P. Abel, M. L. Lykins, Gargi Shaw, W. J. Henney, and P. C. Stancil, 2013, Rev. Mex. Soc., 49, 1.





\bibitem[\protect\astroncite{Furlong et al.}{2015a}]{furlong2015a}
Furlong, M., Bower, R.~G., Theuns, T., et al.\ 2015, MNRAS, 450, 4486   

\bibitem[\protect\astroncite{Furlong et al.}{2015b}]{furlong2015b}
Furlong, M., Bower, R.~G., Crain, R.~A., et al.\ 2015, arXiv:1510.05645

\bibitem[\protect\astroncite{Giallongo et al.}{2015}]{giallongo2015} 
Giallongo, E., Grazian, A., Fiore, F., et al., 2015, AAP, 578, A83

\bibitem[\protect\astroncite{Graham}{2016}]{graham2016} 
Graham, A.~W., 2016, Galactic Bulges, 418, 263 

\bibitem[\protect\astroncite{Grogin et al.}{2011}]{grogin2011} 
Grogin, N. A., Kocevski, D., Faber, S. et al. 2011, ApJS, 197, 37





\bibitem[\protect\astroncite{Haardt \& Madau.}{2001}]{haardt2001} 
{Haardt}, F. and {Madau}, P., 2001, Clusters of Galaxies and the High Redshift Universe Observed in X-rays, 
astro-ph/0106018. 


\bibitem[\protect\astroncite{Hasinger et al.}{2005}]{hasinger2005}
{Hasinger}, G. and {Miyaji}, T. and {Schmidt}, M., 2005, A\&A, 441, 417.

\bibitem[\protect\astroncite{Hasinger}{2008}]{hasinger2008}
{Hasinger}, G., 2008, A\& A, 490, 905.


\bibitem[\protect\astroncite{Hirschmann et al.}{2014}]{hirschmann2014}
Hirschmann, M., Dolag, K., Saro, A., et al.\ 2014, MNRAS, 442, 2304 


\bibitem[\protect\astroncite{Hirschmann et al.}{2012}]{hirschmann2012}
{Hirschmann} M.,  {Somerville} R.~S.,  {Naab} T.,    {Burkert} A.,  2012,
  MNRAS, 426, 237



\bibitem[\protect\astroncite{Hopkins et al.}{2007}]{hopkins2007}  
Hopkins, P.~F.,  Richards, G.~T., \& Hernquist, L., 2007, ApJ, 654, 731 

\bibitem[\protect\astroncite{Hopkins et al.}{2009}]{hopkins2009}  
{Hopkins} P.~F., and {Hernquist} L.,  {Cox} T.~J. , {Keres} D. and 
{Wuyts} S., 2009, ApJ, 691, 1424.


\bibitem[\protect\astroncite{Jenkins}{2010}] {jenkins2010}
Jenkins ~A., 2010, MNRAS, 403, 1859.

\bibitem[\protect\astroncite{Jenkins}{2013}] {jenkins2013}
Jenkins ~A., 2013, MNRAS, 434, 2094.



\bibitem[\protect\astroncite{Kelly \&  Shen}{2013}]{kelly2013}
Kelly B.~C.,  Shen Y.,  2013, ApJ, 764, 45.




\bibitem[\protect\astroncite{Kennicutt}{1998}] {kennicut1998}
Kennicutt, Jr., R. C., 1998, ARA\&A, 36, 189.

\bibitem[\protect\astroncite{Khandai et al.}{2015}]{khandai2015} 
Khandai, N. and Di Matteo, T. and Croft, R. and Wilkins, S.~M. and 
	Feng, Y. and Tucker, E. and DeGraf, C. and Liu, M.-S., 2015,
MNRAS, 450, 1349 

\bibitem[\protect\astroncite{Koekemoer et al.}{2011}]{koekemoer2011} 
Koekemoer, A. M., Faber, S., Ferguson, H. et al. 2011, ApJS, 197, 36



\bibitem[\protect\astroncite{Kormendy \& Ho }{2013}]{kormendy_ho2013} 
Kormendy, J. and Ho, L.~C., 2013, ARA\&A ,51,511.



\bibitem[\protect\astroncite{Lagos et al.}{2015}]{lagos2015}
Lagos, C.~d.~P., Crain, R.~A., Schaye, J., et al., 2015, arXiv:1503.04807

\bibitem[\protect\astroncite{Lansbury et al.}{2015}]{lansbury2015}
Lansbury, G.~B., Gandhi, P., Alexander, D.~M., et al., 2015, ApJ, 809, 115 

\bibitem[\protect\astroncite{Lewis et al.}{2000}]{lewis2000}
Lewis A., Challinor A., Lasenby A., 2000, ApJ, 538, 473





\bibitem[\protect\astroncite{Lusso et al.}{2012}]{lusso2012} 
Lusso, E., Comastri, A., Simmons, B.~D., et al., 2012, MNRAS, 425, 623 

 
\bibitem[\protect\astroncite{Madau \& Haardt.}{2015}]{madau2015} 
Madau, P., \& Haardt, F., 2015, ApJ, 813, L8 


\bibitem[\protect\astroncite{Marconi et al.}{2004}]{marconi2004} 
{Marconi}, A. and {Risaliti}, G. and {Gilli}, R. and {Hunt}, L.~K. and 
{Maiolino}, R. and {Salvati}, M., 2004, MNRAS, 351, 169.   


\bibitem[\protect\astroncite{McAlpine et al.}{2015}]{mcAlpine2015} 
McAlpine S., Helly J.C., Schaller M., et al.\ 2015, arXiv:1510.01320

\bibitem[\protect\astroncite{McCarthy et al.}{2010}]{mcCarthy2010} 
McCarthy  I.~G., Schaye ~J., Ponman T.~J., Bower R.~G., Booth C.~M., Dalla Vecchia C., Crain R.~A. and Springel V.,
2010, MNRAS,  406, 822.


\bibitem[\protect\astroncite{McConnell \& Ma}{2013}]{mcConnell_ma2013} 
McConnell, N.~J., \& Ma, C.-P., 2013, ApJ, 764, 184 



\bibitem[\protect\astroncite{Miyaji et al.}{2000}]{miyaji2000}
{Miyaji} T.,  {Hasinger} G., {Schmidt} M.,  2000, A\&A, 353, 25.


\bibitem[\protect\astroncite{Miyaji et al.}{2015}]{miyaji2015}
Miyaji, T., Hasinger, G., Salvato, M., et al., 2015, ApJ, 804, 104


\bibitem[\protect\astroncite{Narayan \& Yi}{1994}]{narayan_yi1994}
{Narayan}, R. and {Yi}, I., 1994,  ApJ, 428, L13.



\bibitem[\protect\astroncite{Planck collaboration et al.}{2013}]{planck13} 
{Planck Collaboration} and {Ade}, P.~A.~R. and {Aghanim}, N. and 
	{Alves}, M.~I.~R. and {Armitage-Caplan}, C. and {Arnaud}, M. and 
	{Ashdown}, M. and {Atrio-Barandela}, F. and {Aumont}, J. and 
	{Aussel}, H. and et al., 2013, arxiv e-prints, 1303.5062.

\bibitem[\protect\astroncite{Portinari et al.}{1998}]{portinari1998} 
Portinari L., Chiosi C., \& Bressan A.\ 1998, A\& A, 334, 505.  


\bibitem[\protect\astroncite{Power et al.}{2010}]{power2011} Power, C. and
Nayakshin, S. and King, A., 2011, MNRAS, 412, 269.


\bibitem[\protect\astroncite{Rees et al.}{1982}]{rees1982}
Rees, M.~J. and Begelman, M.~C. and Blandford, R.~D. and 
Phinney, E.~S., 1982, NATURE, 295, 17.


\bibitem[\protect\astroncite{Rosas-Guevara et al.}{2015}]{rosas-guevara2015}
Rosas-Guevara Y. M., Bower R.G., Schaye J., et al.\ 2015, MNRAS, 454, 1038 



\bibitem[\protect\astroncite{Shakura \& Syunyaev}{1973}]{shak73} 
Shakura N.I ., Syunyaev R. A., 1973, A\&A , 24, 337.

\bibitem[\protect\astroncite{Schaller et al.}{2015}]{schaller2015}
Schaller, M., Dalla Vecchia, C., Schaye, J., et al., 2015, MNRAS, 454, 2277


\bibitem[\protect\astroncite{Schaye}{2004}]{schaye2004} 
Schaye, J., 2004, ApJ, 609, 667 


\bibitem[\protect\astroncite{Schaye \& Dalla Vecchia}{2008}]{scha08} 
Schaye J., Dalla Vechia C., 2008, MNRAS, 383, 1210.

\bibitem[\protect\astroncite{Schaye et al.}{2010}]{scha10} 
Schaye ~J., Dalla Vecchia ~C., Booth C.~M., Wiersma R.~P.~C., Theuns ~T., Haas M.~R., Bertone ~S. and Duffy A.~R.{\it et al.},
 2010, MNRAS, 402, 1536. 


\bibitem[\protect\citeauthoryear{Schaye et al.}{2015}]{schaye2015} 
Schaye, J., Crain, R.~A., Bower, R.~G., et al.\ 2015, MNRAS, 446, 521 

\bibitem[\protect\astroncite{Schmidt}{1959}]{schmidt1959} 
Schmidt, M., 1959, ApJ, 129, 243.

 

\bibitem[\protect\astroncite{Shankar et al.}{2004}]{shankar2004} 
{Shankar}, F. and {Salucci}, P. and {Granato}, G.~L. and {De Zotti}, G. and 
{Danese}, L., 2004, MNRAS, 354, 1020.


\bibitem[\protect\astroncite{Shankar et al.}{2013}]{shankar2013} 
Shankar, F., 2013, Classical and Quantum Gravity, 30, 244001

\bibitem[\protect\astroncite{Sijacki et al.}{2007}]{sijacki2007} 
Sijacki ~D., Springel ~V., Di Matteo ~T. and Hernquist ~L., 2007
MNRAS,  380, 877.

\bibitem[\protect\astroncite{Sijacki et al.}{2015}]{sijacki2015} 
Sijacki, D., Vogelsberger, M., Genel, S., et al.\ 2015, MNRAS, 452, 575 


\bibitem[\protect\astroncite{Soltan}{1982}]{soltan1982} 
Soltan ~A., 1982, MNRAS, 200, 115.




\bibitem[\protect\astroncite{Springel et al.}{2005a}]{springel2005a}
Springel ~V., Di Matteo ~T. and Hernquist ~L., 2005,
MNRAS, 361, 776.

\bibitem[\protect\astroncite{Springel}{2005b}]{springel2005b}
Springel, ~V., 2005b, MNRAS, 364, 1105.

\bibitem[\protect\astroncite{Steffen et al.}{2003}]{steffen2003}
{Steffen}, A.~T. and {Barger}, A.~J. and {Cowie}, L.~L. and 
	{Mushotzky}, R.~F. and {Yang}, Y., 2003, ApJ, 596, L23.


\bibitem[\protect\astroncite{Trayford et al.}{2016}]{trayford2016} 
Trayford, J.~W., Theuns, T., Bower, R.~G., et al.\ 2016, arXiv:1601.07907 

	

\bibitem[\protect\astroncite{Tremaine et al.}{2002}]{tremaine2002} 
Tremaine ~S., Gebhardt ~K., Bender ~R., Bower ~G., Dressler ~A., Faber S. ~M., Filippenko A. ~V. and Green ~R. {\it et al.},
2002, ApJ, 574, 740.


\bibitem[\protect\astroncite{Treister et al.}{2009}]{treister2009} 
Treister, E., Urry, C. M., \& Virani, S., 2009, ApJ, 696, 110

\bibitem[\protect\astroncite{Ueda et al.}{2003}]{ueda2003} 
{Ueda}, Y. and {Akiyama}, M. and {Ohta}, K. and {Miyaji}, T.,
2003,ApJ , 598, 886.

\bibitem[\protect\astroncite{Ueda et al.}{2014}]{ueda2014}
 Ueda, Y., Akiyama, M., Hasinger, G., Miyaji, T., \& Watson, M.~G.\ 2014, ApJ, 786, 104 

 





\bibitem[\protect\astroncite{Vasudevan \& Fabian}{2009}]{vasudevan2009} 
Vasudevan R. V., Fabian A. C., 2009, MNRAS, 392, 1124

\bibitem[\protect\astroncite{Vogelsberger et al.}{2014}]{vogelsberger2014} 
Vogelsberger, M., Genel, S., Springel, V., et al., 2014, MNRAS, 444, 1518




\bibitem[\protect\astroncite{ Yu \& Tremaine}{2002}]{yu_tremaine_2002}
{Yu}, Q. and {Tremaine}, S., 2002, MNRAS, 335, 965.



\bibitem[\protect\astroncite{Wiersma et al.}{2009a}]{wiersma2009a}
Wiersma R.~P.~C., Schaye ~J., Theuns ~T., Smith B.~D,
2009,  MNRAS, 393, 99.

\bibitem[\protect\astroncite{Wiersma et al.}{2009b}]{wiersma2009b}
Wiersma R.~P.~C., Schaye ~J., Theuns ~T., Dalla Vecchia ~C. and Tornatore ~L.,
2009,  MNRAS, 399, 574.


\end{thebibliography}
\end{document}